\documentclass{mn2e}
\usepackage{epsfig}

\usepackage{graphicx}

\newif\ifAMStwofonts
\AMStwofontstrue

\title[SHADES paper IX]
{
The SCUBA Half Degree Extragalactic Survey (SHADES) - IX: 
the environment, mass and redshift dependence of star formation 
}
\author[Serjeant et al.]
  {
{S. Serjeant$^{1}$, S. Dye$^2$, A. Mortier$^3$, J. Peacock$^3$, E. Egami$^4$, 
  M. Cirasuolo$^3$, G. Rieke$^4$, }\vspace*{0.3cm}\\  
{\LARGE\rm  C. Borys$^5$, S. Chapman$^6$, 
D. Clements$^7$, K. Coppin$^8$, J. Dunlop$^3$, S. Eales$^2$,
  D. Farrah$^9$,}\vspace*{0.3cm}\\
{\LARGE\rm  M. Halpern$^{10}$, P. Mauskopf$^2$, A. Pope$^{10,11}$,
M. Rowan-Robinson$^7$, D. Scott$^{10}$,}\vspace*{0.3cm}\\
{\LARGE\rm I. Smail$^8$, M. Vaccari$^{12}$}\vspace*{0.3cm}\\
$^1$Department of Physics and Astronomy, The Open University, Milton Keynes, MK7 6AA, UK\\
$^2$Cardiff University, School of Physics \& Astronomy, Queens
Buildings, The Parade, Cardiff, CF24 3AA, UK\\
$^3$Institute for Astronomy, The University of Edinburgh, Royal
Observatory, Blackford Hill, Edinburgh EH9 3HJ, UK\\
$^4$Department of Astronomy, University of Arizona, 933 N Cherry
Avenue, Rm. N204, Tucson, Arizona 85721-0065, USA\\
$^5$University of Toronto, Department of Astronomy \& Astrophysics, 60
St. George Street, Toronto, Ontario, Canada, M5S 3H4, Canada\\
$^6$Institute of Astronomy, University of Cambridge, Madingley Road,
Cambridge, CB3 0HA, UK\\
$^7$Astrophysics Group, Imperial College London, Blackett Laboratory,
Prince Consort Road, London, SW7 2AZ, UK\\
$^8$Institute for Computational Cosmology, Durham University, South
Road, Durham, DH1 3LE, UK\\
$^9$106 Space Sciences Building, Cornell University, Ithaca, NY 14853,
USA\\ 
$^{10}$Department of Physics \& Astronomy, University of British
Columbia, 6224 Agricultural Road, Vancouver, B.C. V6T 1Z1, Canada\\ 
$^{11}$ {\it Spitzer} Fellow; National Optical Astronomy Observatory,
950 North Cherry Avenue, Tucson, AZ 85719, U.S.A.\\ 
$^{12}$Department of Astronomy, University of Padova,
vic. Osservatorio 2, 35122, Padova - I, Italy\\ 
}
\date{Received 2007}
\pubyear{2007}

\begin{document}

\input BoxedEPS.tex
\SetEPSFDirectory{./}

\SetRokickiEPSFSpecial
\HideDisplacementBoxes

\label{firstpage}

\maketitle

\begin{abstract}
We present a comparison between the SCUBA Half Degree Extragalactic
Survey (SHADES) at $450\,\mu$m and $850\,\mu$m in the Lockman Hole
East with a deep Spitzer Space Telescope survey at $3.6-24\,\mu$m
conducted in Guaranteed Time.  Using stacking analyses we 
demonstrate a striking correspondence between the galaxies 
contributing the submm extragalactic
background light, with those likely to dominate the backgrounds at
Spitzer wavelengths.  Using a combination $BRIzK$ plus Spitzer
photometric redshifts, we show that at least a third of the
Spitzer-identified submm galaxies at $1<z<1.5$ appear to reside in
overdensities when the density field is smoothed at $0.5-2\,$Mpc
comoving diameters, supporting the high-redshift reversal of the local
star formation -- galaxy density relation.
We derive the dust-shrouded
cosmic star formation history of galaxies as a function of assembled 
stellar masses. 
For model stellar masses $<10^{11}M_\odot$, this peaks at lower redshifts
than the ostensible $z\sim2.2$ maximum for submm point sources, adding
to the growing consensus for ``downsizing'' in star formation. Our 
surveys are also consistent with ``downsizing'' in mass assembly. 
Both the mean star formation rates $\langle dM_*/dt\rangle$ and specific star 
formation rates $\langle (1/M_*)dM_{*}/dt\rangle$ are in striking disagreement with 
some semi-analytic predictions from the Millenium simulation. 
The discrepancy could either be resolved with a top-heavy initial
mass function, or a significant component of the submm flux heated
by the interstellar radiation field.
\end{abstract}


\section{Introduction}\label{sec:introduction}
The SCUBA Half Degree Extragalactic Survey (SHADES, Mortier et
al. 2005, Coppin et al. 2006) is a long-term submm survey conducted at
the James Clerk Maxwell Telescope from 2003-2005.  A key goal of
SHADES has been to determine whether submm galaxies are the likely
progenitors of giant ellipticals. The clustering of submm galaxies is
a strong discriminant of competing models (van Kampen et al. 2005),
and measurements of the angular correlation function of submm galaxies
in broad $\Delta z\simeq0.5$ redshift shells is one of the principal
experimental aims of SHADES. In this paper we will take a different
approach to the problem, by estimating the matter overdensities in
which submm galaxies reside {\it via} the assembled stellar masses in
the submm galaxy environments. The SHADES survey was conducted in two
fields, each with abundant multi-wavelength supporting survey
data. The $\sim0.1$ deg$^2$ surveyed by SHADES in the Lockman Hole
East field, in particular, has some of the best Spitzer Space
Telescope data of any contiguous field over hundreds of square
arcminutes. The comparison between SHADES and this Spitzer data, which
was taken in Spitzer guaranteed time, forms the basis of our
constraints on the submm galaxy environments, and allows us important
new insights on the submm extragalactic background light.

\begin{figure}
  \ForceWidth{5.0in}
  \hSlide{-1cm}
  \BoxedEPSF{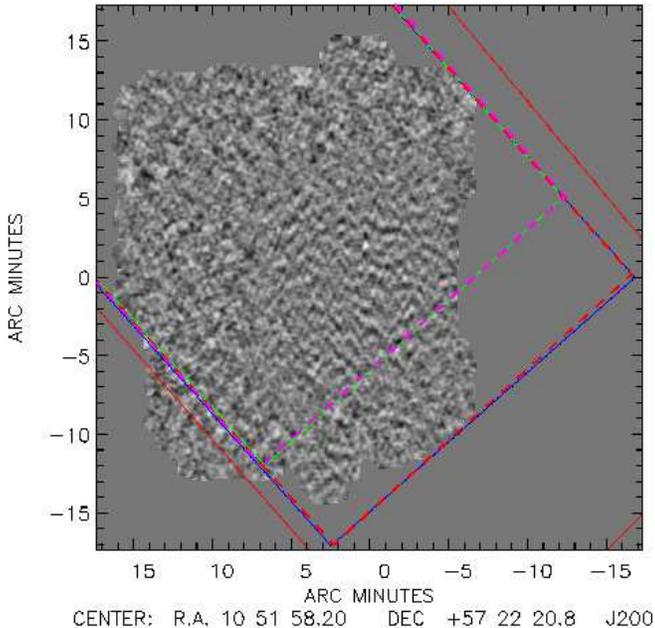}
\caption{\label{fig:coverage} 
The SHADES $850\,\mu$m map with point sources subtracted, used for stacking
analyses in this paper (greyscale). Also plotted is the coverage of
the Spitzer $24\,\mu$m data (solid red line), the $3.6\,\mu$m data (solid
blue line), $4.5\,\mu$m (solid green line), $5.8\,\mu$m (dashed red line),
and $8\,\mu$m (dashed magenta line). Note the availability of Spitzer
data over most of the SHADES area. 
}
\end{figure}

The galaxies that dominate the extragalactic background light at any 
given redshift are
necessarily the same as those which dominate the comoving
volume-averaged luminosity density at that redshift 
(e.g. Peacock 1999). The favourable
K-corrections in the submm make the submm extragalactic background
light sensitive to the cosmic star formation history throughout most
of the history of the Universe. Resolved submm point sources from
blank field surveys (i.e. those at the few-mJy level) 
contribute a few tens of percent to the $850\,\mu$m extragalactic
background, but cannot account for all of it (Hughes et al. 1998,
Barger et al. 1998, 1999, 
Blain et al. 1999, Eales et al. 2000, Scott et al. 2002, Smail et
al. 2002, Cowie et al. 2002, Scott et al. 2006). At
$350-450\,\mu$m there are very few reliably detected resolved point
sources (e.g. Scott et al. 2002, Khan et al. 2005,
2007), and those that have been detected are far from accounting for
the majority of the $450\,\mu$m extragalactic background light. 
However, $850\,\mu$m-selected galaxies can be readily detected at 
$350-450\,\mu$m and can account for a minority of the $350-450\,\mu$m 
background (e.g. Chapman et al. 2005, Khan et al. 2005, 2007, 
Kov\'{a}cs et al. 2006, Coppin et al. 2007). 

\begin{figure}
  \ForceWidth{4in}
  \hSlide{-1cm}
  \BoxedEPSF{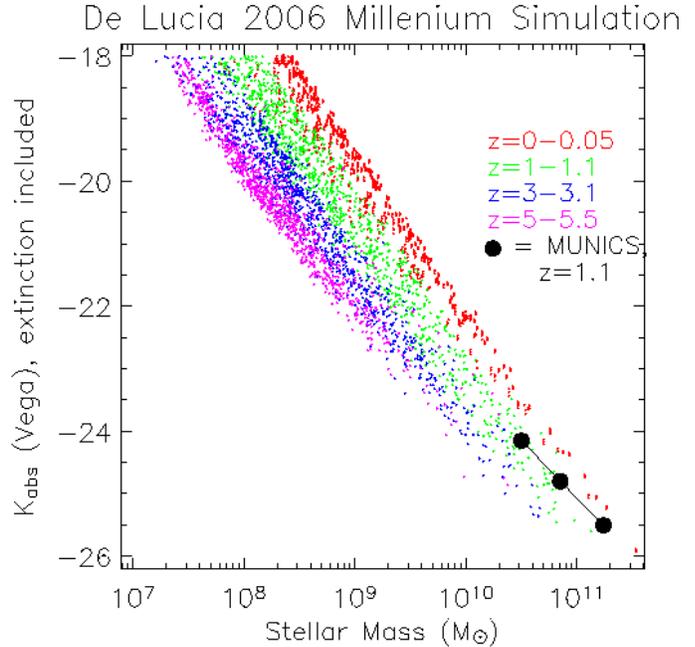}
\caption{\label{fig:millenium_conversion} Correlation between model
  stellar mass and absolute K magnitude predicted using the Millenium
  simulation by de Lucia (2006), for several redshift shells. Also
  plotted are the data from the MUNICS survey (Drory et
  al. 2004). Note the small dispersion and the clear redshift
  dependence.}
\end{figure}

\begin{figure}
  \ForceWidth{4.5in}
  \hSlide{-1cm}
  \BoxedEPSF{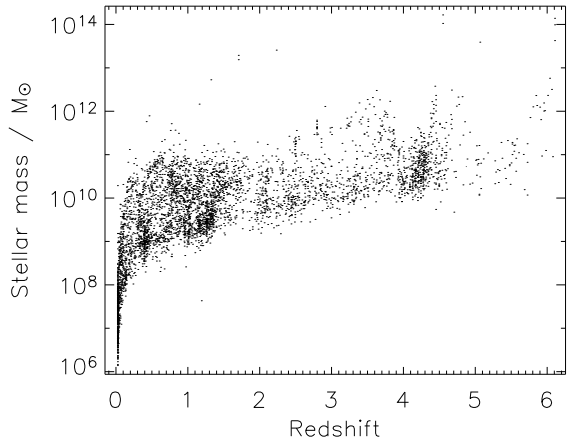}
\caption{\label{fig:mstar_vs_z} Model stellar mass estimates of
  Spitzer galaxies discussed in the text, versus photometric redshifts
  from Dye et al. (2008, SHADES paper VII). Note that the model
  stellar masses above $z=0.5$ depend only weakly on photometric
  redshift.}
\end{figure}

\begin{figure}
  \ForceWidth{5.0in}
  \hSlide{-2cm}
  \BoxedEPSF{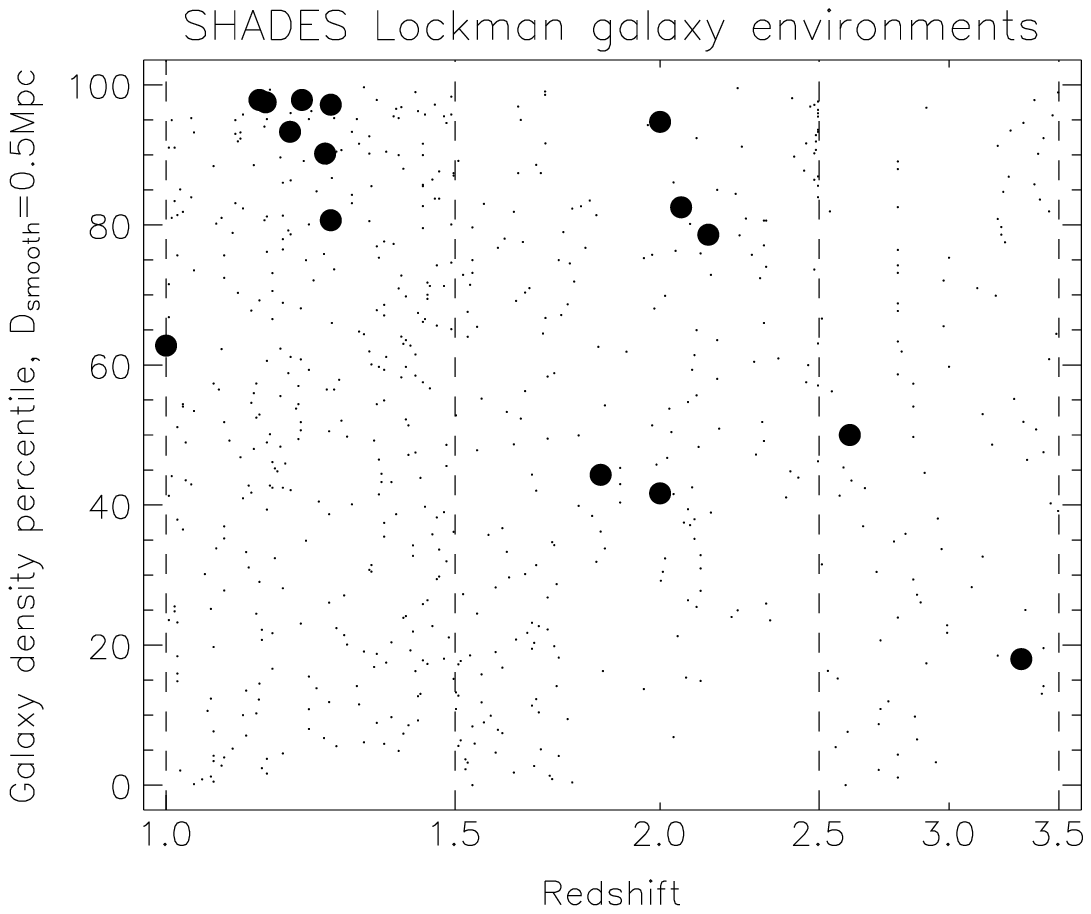}
\caption{\label{fig:percentiles} Density distribution percentiles for
  submm galaxies (large symbols), compared with that of
  $3.6\,\mu$m-selected galaxies (dots). The matter density
  distribution estimated from Spitzer $3.6\,\mu$m imaging in broad
  redshift bins (indicated by dashed lines) was smoothed with a $0.5$
  comoving Mpc diameter top hat kernel as discussed in the
  text. Regions with zero estimated density were excluded from this
  test. Note that submm galaxies sample the highest $10-20$
  percentiles of this matter density distribution at lower redshifts.}
\end{figure}

\begin{figure}
  \ForceWidth{5.0in}
  \hSlide{-2cm}
  \BoxedEPSF{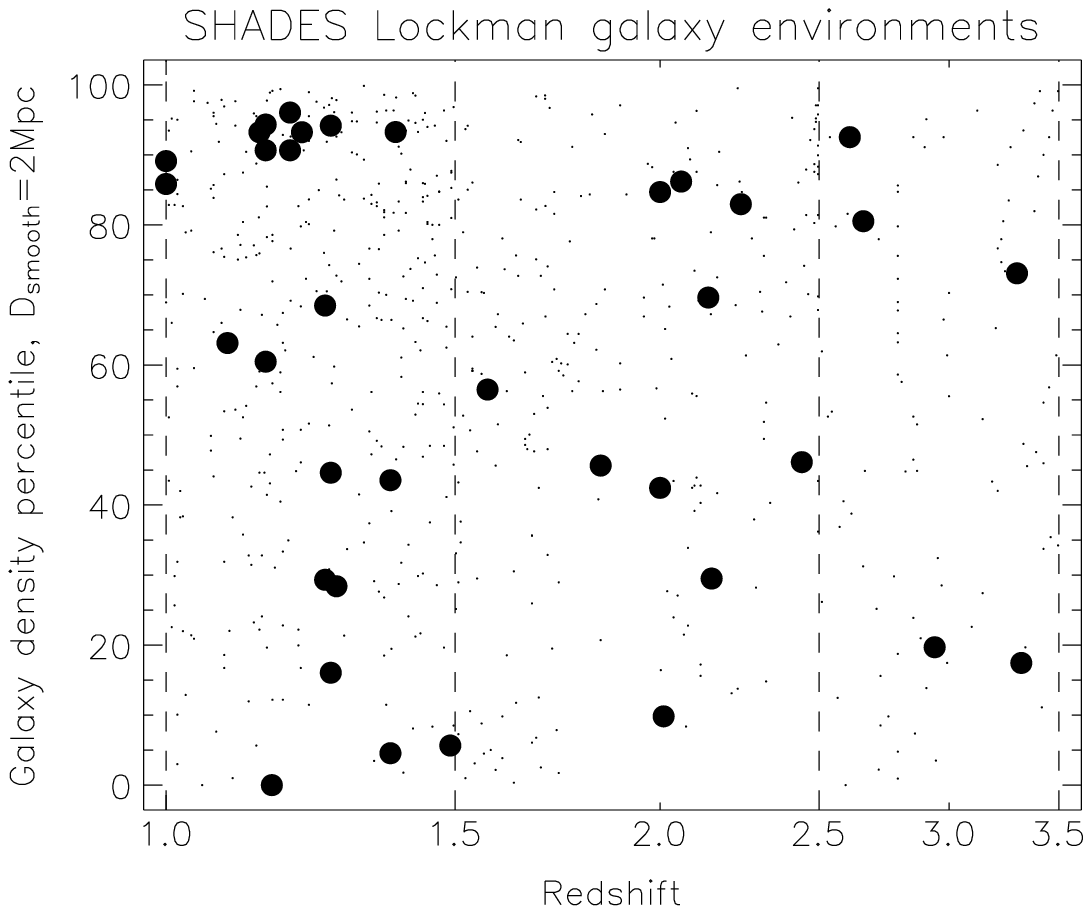}
\caption{\label{fig:percentiles_2Mpc} Density distribution percentiles
  for submm galaxies, compared with that of $3.6\,\mu$m-selected
  galaxies (dots). The matter density distribution estimated from
  Spitzer $3.6\,\mu$m imaging in broad redshift bins (indicated by
  dashed lines) was smoothed with an $2$ comoving Mpc diameter top hat
  kernel as discussed in the text. Regions with zero estimated density
  were excluded from this test. Note that around half of the submm
  galaxies in the lower redshift bin sample the highest $10-20$
  percentiles of this matter density distribution, when smoothed to
  this scale. The larger smoothing width reduces the area with no
  density field data, so the sample of submm galaxies is larger.}
\end{figure}

\begin{table}
\begin{tabular}{|l|llllll}
   z & 0.025  & 1.05 & 2.05 & 3.05 & 4.1 & 5.25 \\
 \hline 
   a & 2.007 & 1.457 & 1.156 & 0.8193 & 0.3064 & 0.5831\\
\hline 
   b & -0.3944 & -0.4010 & -0.4058 & -0.4178 & -0.4373 & -0.4190\\
\end{tabular}
\caption{\label{tab:kfit}Best-fit values for the conversion between
  K-band AB absolute magnitude and stellar mass using equation
  \ref{eqn:kmconv}, from the de Lucia (2005) Millenium simulation. The
  top row gives the redshift, and the other rows give the fitted
  parameters in a narrow redshift bin centred on that redshift.}
\end{table}

This situation, particularly at $450\,\mu$m, will change with the advent
of the SCUBA-2 camera on the James Clerk Maxwell Telescope. In the
meantime, attention has focussed on stacking analyses. Instead of
aiming to detect individual resolved galaxies, this approach seeks to
detect the average signal from a population. This has met with some
success. Peacock et al. 2000 found a $\sim3\sigma$ signal at $850\,\mu$m
from the Lyman-break population in the Hubble Deep Field North. The
submm:UV flux ratio suggested an obscuration very different to that of
the submm point source population, and there were hints of a flat
redshift distribution at $z>1$ in these faint submm-emitting galaxies. 
Further stacking analyses of extremely red galaxies (e.g. Webb et
al. 2004, Takagi et al. 2007) found them to contribute a significant
minority of the obscured star formation history. 
A submm stacking analysis of near-infrared and mid-infrared
selected galaxies from the $5'\times5'$ Spitzer Early Release
Observations (Serjeant et al. 2004) found that Spitzer $5.8\,\mu$m and
$8\,\mu$m populations could account for around a quarter of the
$850\,\mu$m extragalactic background light, and the majority of the
$450\,\mu$m background, albeit in a small sample. The sample size was
not large enough in this study to distinguish the stacked signal from
low-redshift red dusty galaxies, and that from high-redshift galaxies.

To constrain the redshift ranges responsible for the submm
extragalactic background light, larger samples were needed. Wang et
al. (2006) and Dye et al. (2006) both made stacking analyses of
Spitzer-selected galaxies, though in the former case it was 
combined with H-band selection. 
Both groups 
found that Spitzer galaxies contribute significantly to the submm
extragalactic background light.
However, the redshift ranges responsible in
these surveys differed, with Wang et al. finding the $z<1$ population
dominating (their figure 12), while Dye et al. found $z>1$ the more
important (their figure 6) though with slightly larger errors. 
One possible explanation for this difference is cosmic variance; 
the wide-area SHADES would be ideal to resolve this controversy. 
Another possibility is the effect of redshifted PAH features in the 
Dye et al. analysis, which would not be present in the $\leq3.6\mu$m 
Wang et al. analysis. 

In this paper we extend these results to a deeper Spitzer catalogue,
and a wider-area submm survey. In a confusion-limited submm survey,
the stacking signal-to-noise is roughly proportional to the square
root of the number of submm beams, and since SHADES is the widest-area
contiguous submm survey to a depth approaching the effective point
source extraction limit (Scott et al. 2002), this is the best
opportunity to date to examine the submm stacking signal of Spitzer
galaxies.

This paper is the ninth in the SHADES series of papers. Paper I
(Mortier et al. 2005) presented the survey design, motivation and data
analysis. Paper II (Coppin et al. 2006) presented further data
analysis, the source counts, the catalogues and the maps. Paper III
(Ivison et al. 2007) gave the radio and Spitzer $24\,\mu$m
identifications of the submm galaxies in SHADES. Paper IV (Aretxaga et
al. 2007) made photometric redshift estimates of the SHADES catalogue
galaxies using the far-infrared to radio spectral energy
distributions. Paper V (Takagi et al. 2007) examined the submm
properties of near-infrared galaxies in the SHADES suvey data in the
Subaru-XMM Deep Field. Paper VI (Coppin et al. 2007) presented the
results of $350\,\mu$m observations of a subset of SHADES
sources. Paper VII (Dye et al. 2008) made fits to the spectral energy
distributions of the SHADES galaxies in the Lockman Hole, and paper
VIII (Clements et al. 2007) performed a similar analysis for the
SHADES galaxies in the Subaru-XMM Deep Field. Paper X (van Kampen et
al. 2007) measures the clustering of the submm galaxies in the SHADES
survey. A further series of papers will concern the $1.1\,$mm data
taken to supplement the SHADES survey with the AzTEC instrument on the
James Clerk Maxwell Telescope.

This paper is structured as follows. The Spitzer and submm data
are summarized briefly in section \ref{sec:data}.  Section
\ref{sec:method} describes our methodology and results.
We 
discuss the context of our results in section \ref{sec:discussion},
and we draw conclusions in section \ref{sec:conclusions}. Throughout
the paper we assume a ``concordance'' cosmology, with density
parameters $\Omega_M=0.3$ and $\Omega_\Lambda=0.7$, and a Hubble
constant of $H_0=72$ km s$^{-1}$ Mpc$^{-1}$.

\begin{figure*}
  \ForceHeight{2.5in}\hSlide{-2.5in}\BoxedEPSF{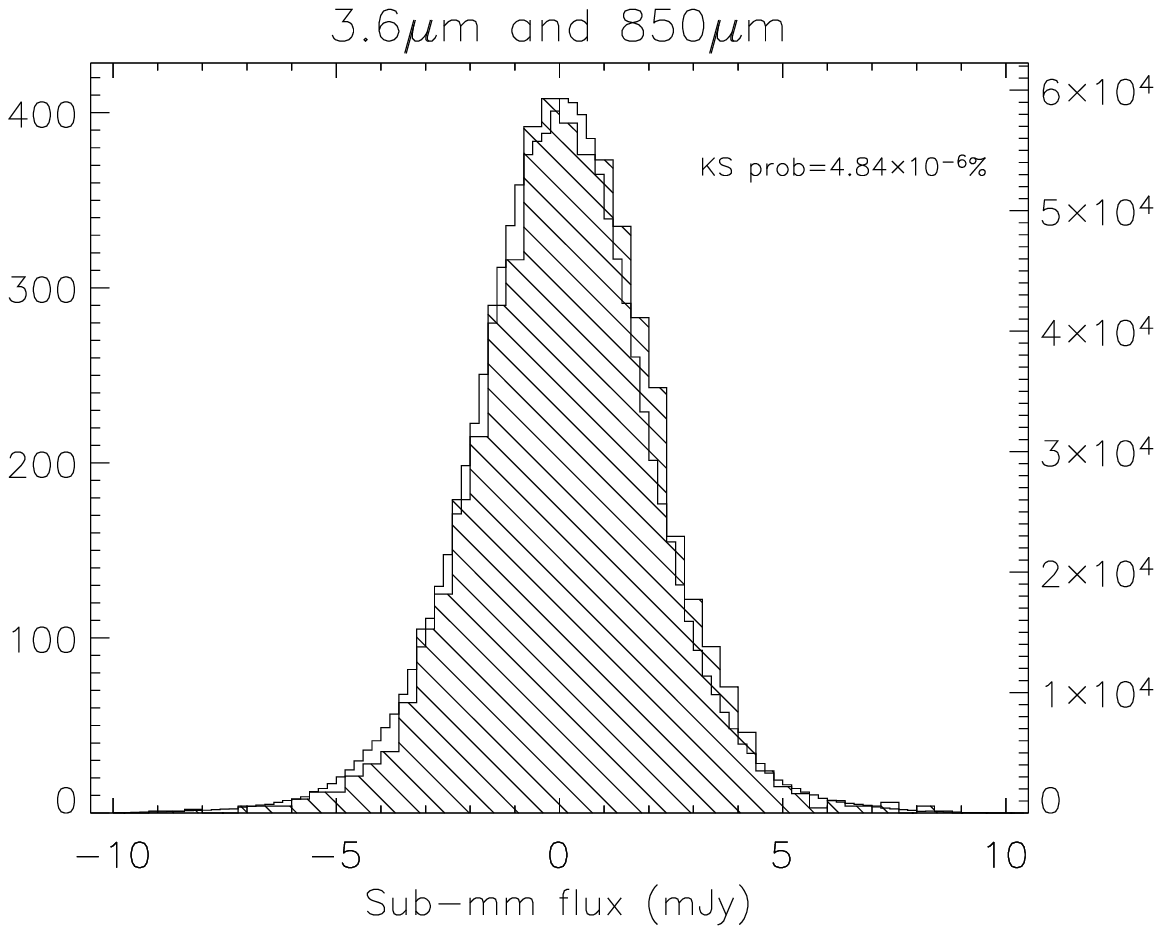}
  \vspace*{-2.5in}\ForceHeight{2.5in}\BoxedEPSF{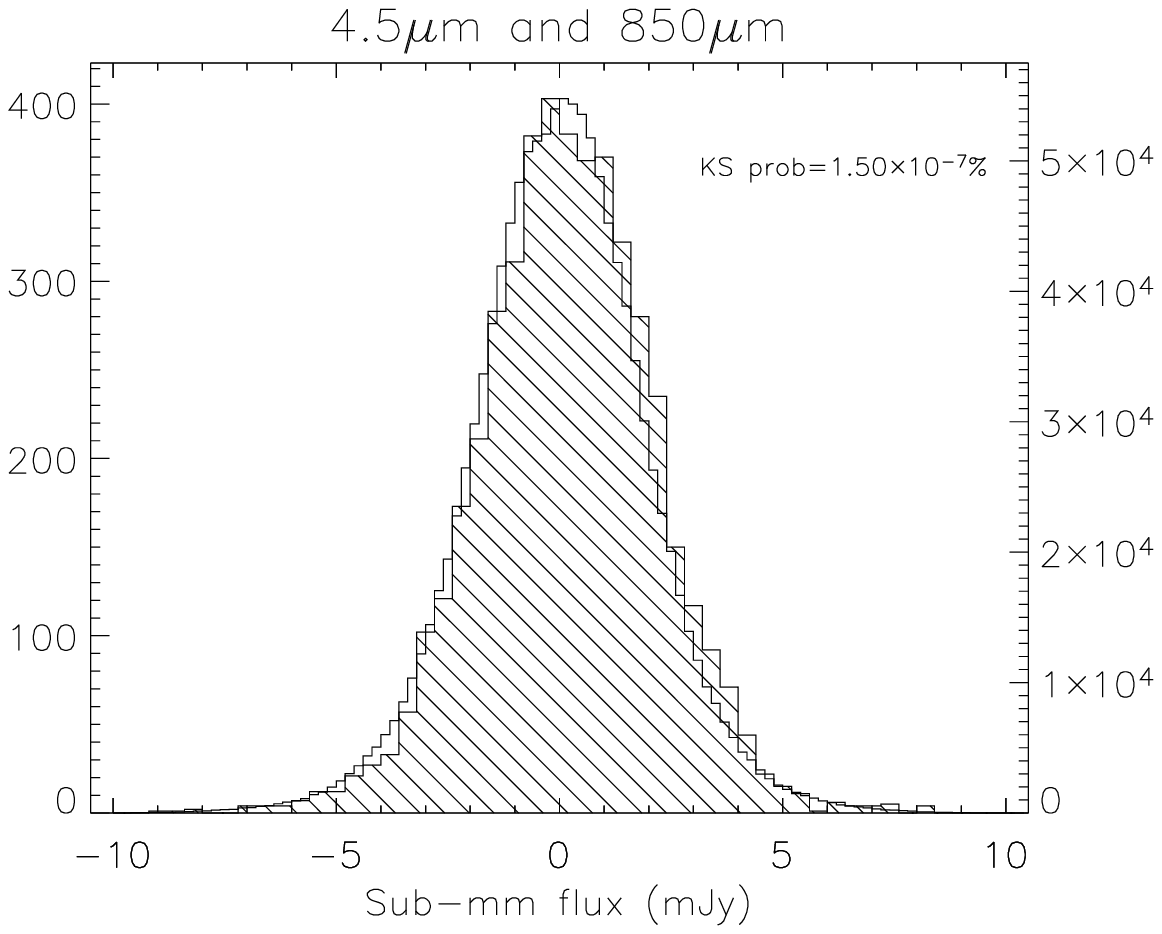}
  \vspace*{-2.5in}\hSlide{2.5in}\ForceHeight{2.5in}\BoxedEPSF{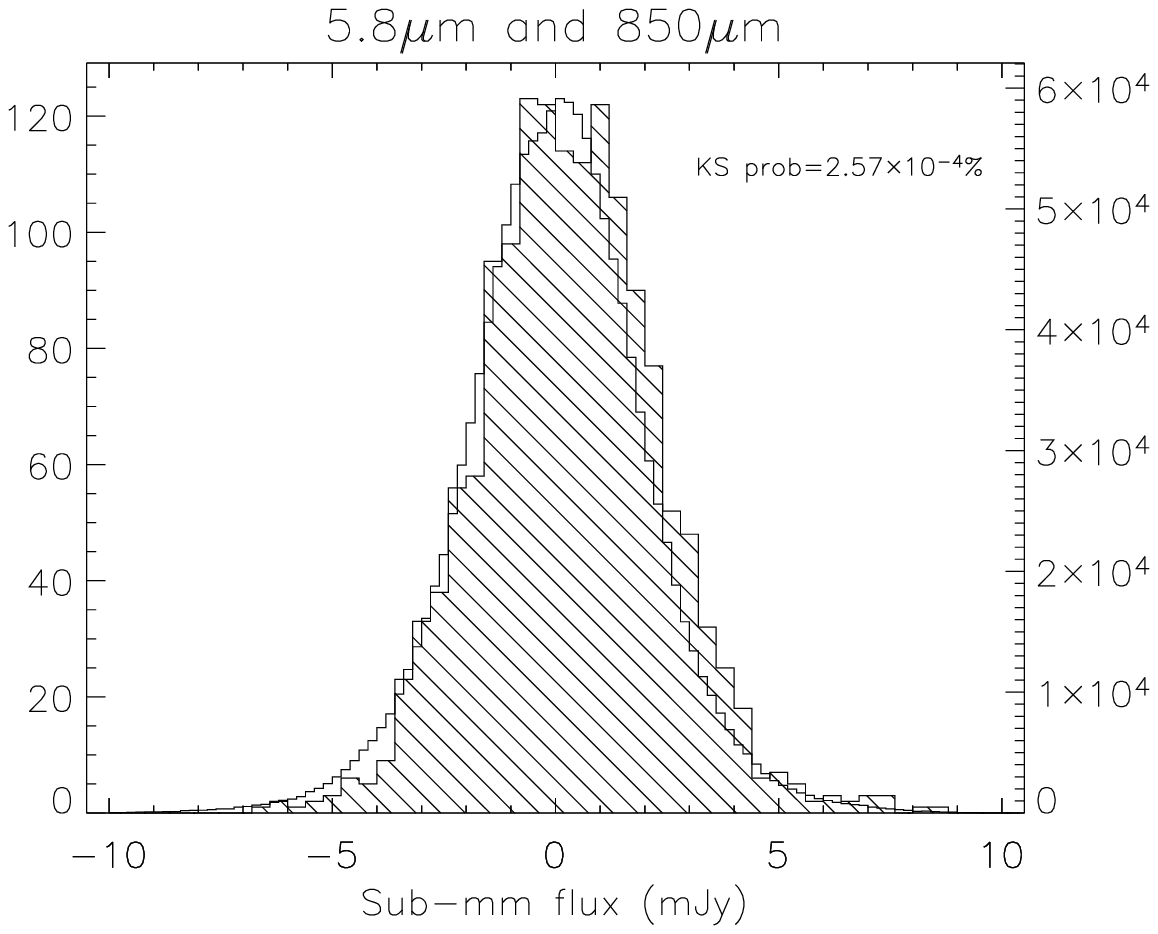}
  \ForceHeight{2.5in}\hSlide{-2.5in}\BoxedEPSF{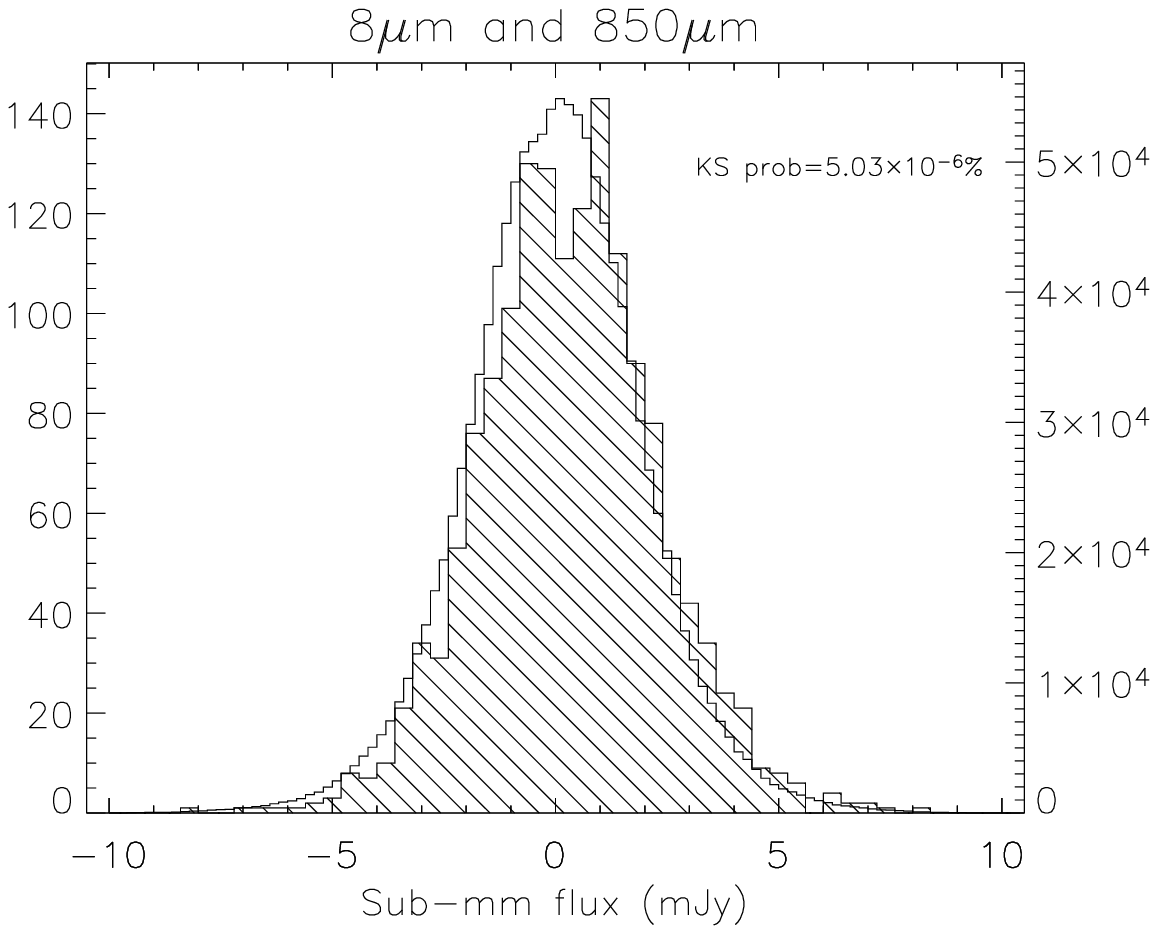}
  \vspace*{-2.5in}\ForceHeight{2.5in}\BoxedEPSF{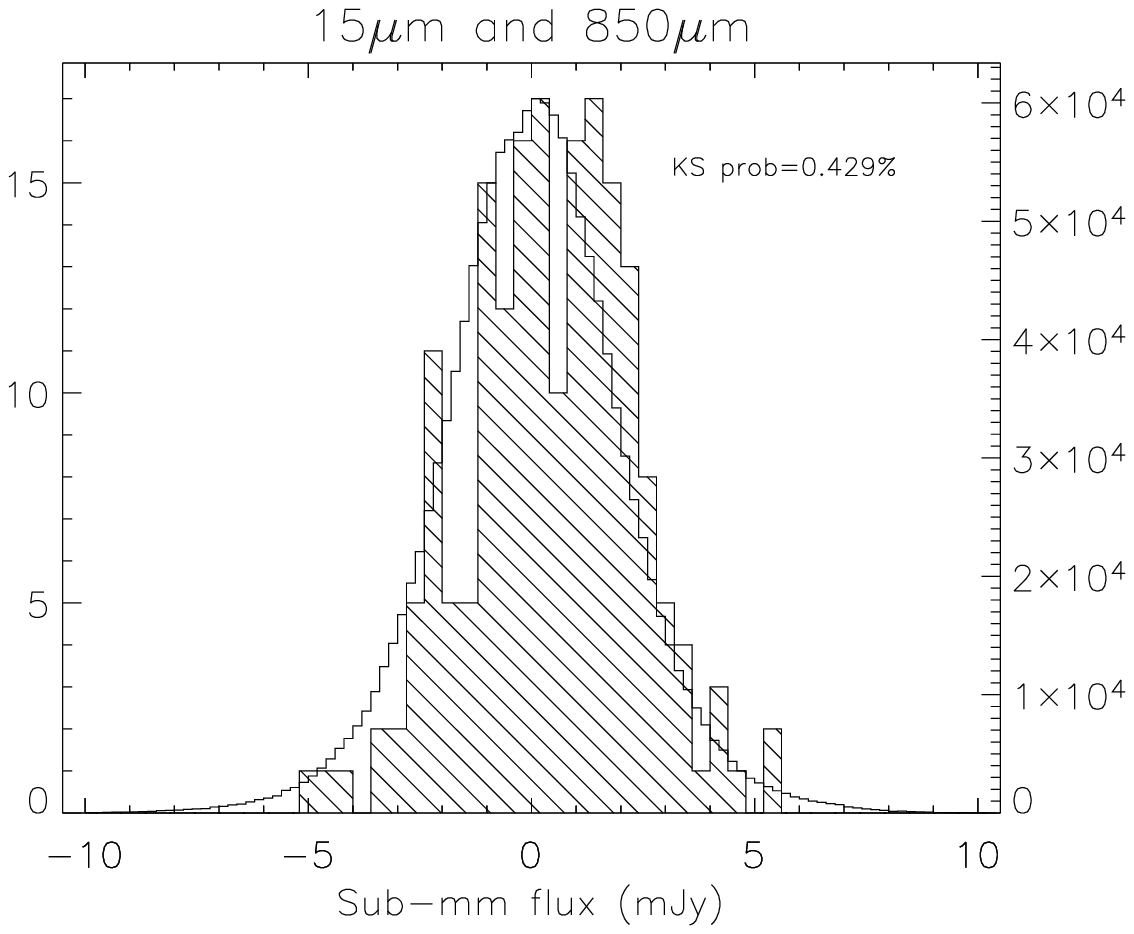}
  \vspace*{-2.5in}\hSlide{2.5in}\ForceHeight{2.5in}\BoxedEPSF{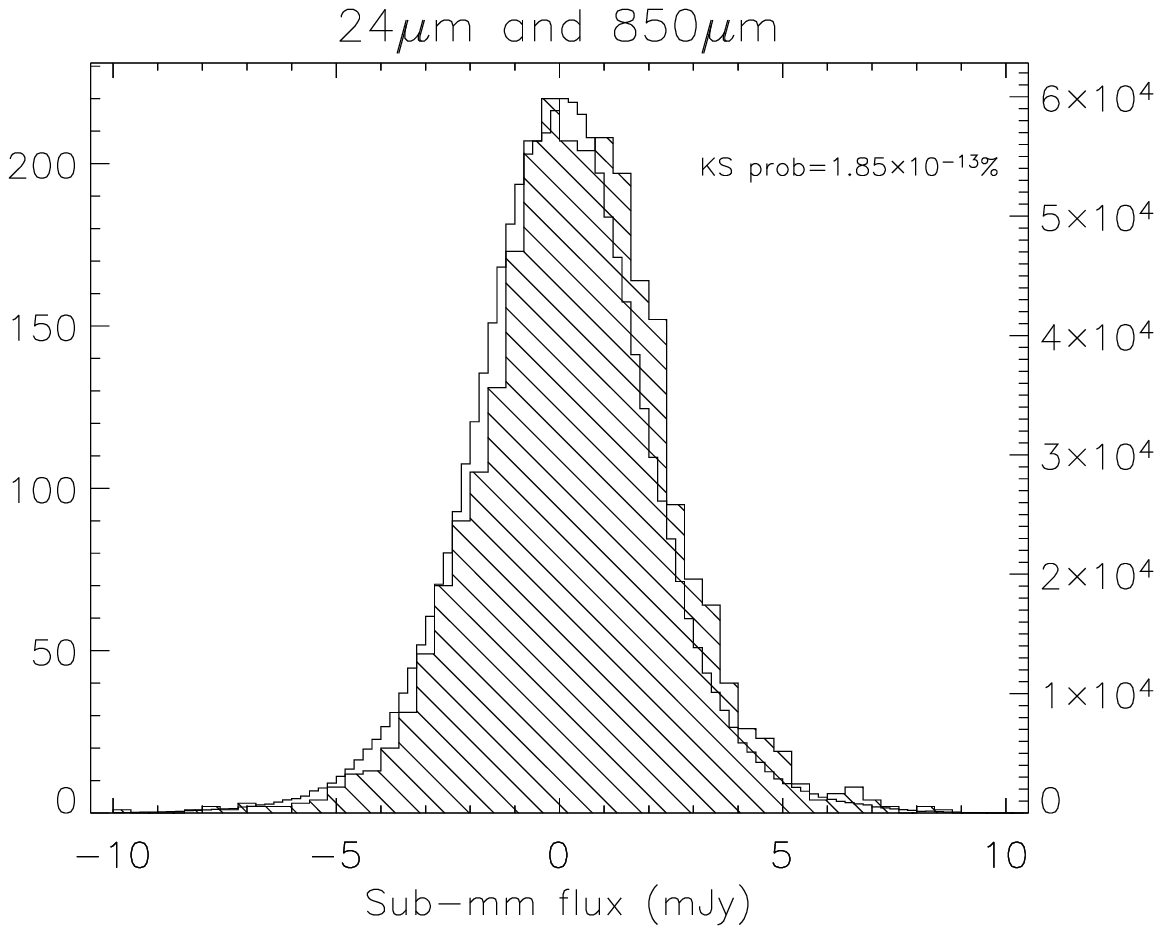}
\caption{\label{fig:850stacks}Histograms of $850\,\mu$m flux in the
  SHADES Lockman map as a whole (unhatched bars) compared to the
  $850\,\mu$m fluxes at the positions of the Spitzer/ISO galaxies selected
  at the indidcated wavelengths (hatched bars). The left-hand vertical scales 
  show the numbers for the hatched histograms, and the right-hand scales refer 
  to the unhatched histograms. Note that in all
  plots, the hatched bars lie typically to the right of the unhatched
  bars. The results of a comparison using the Kolmogorov-Smirnoff test
  are also shown in the diagrams. Submm point sources have been removed.}
\end{figure*}
\begin{figure*}
  \ForceHeight{2.5in}\hSlide{-2.5in}\BoxedEPSF{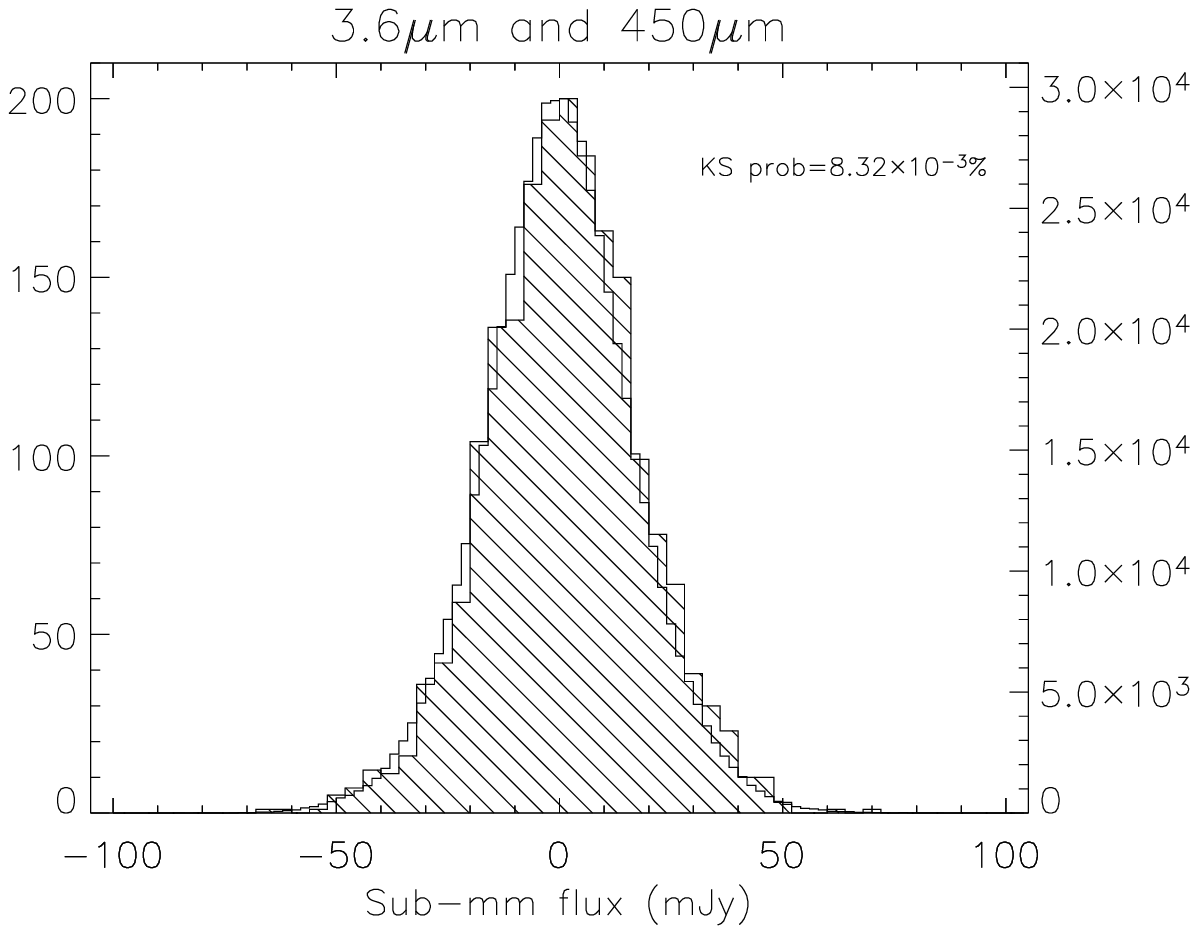}
  \vspace*{-2.5in}\ForceHeight{2.5in}\BoxedEPSF{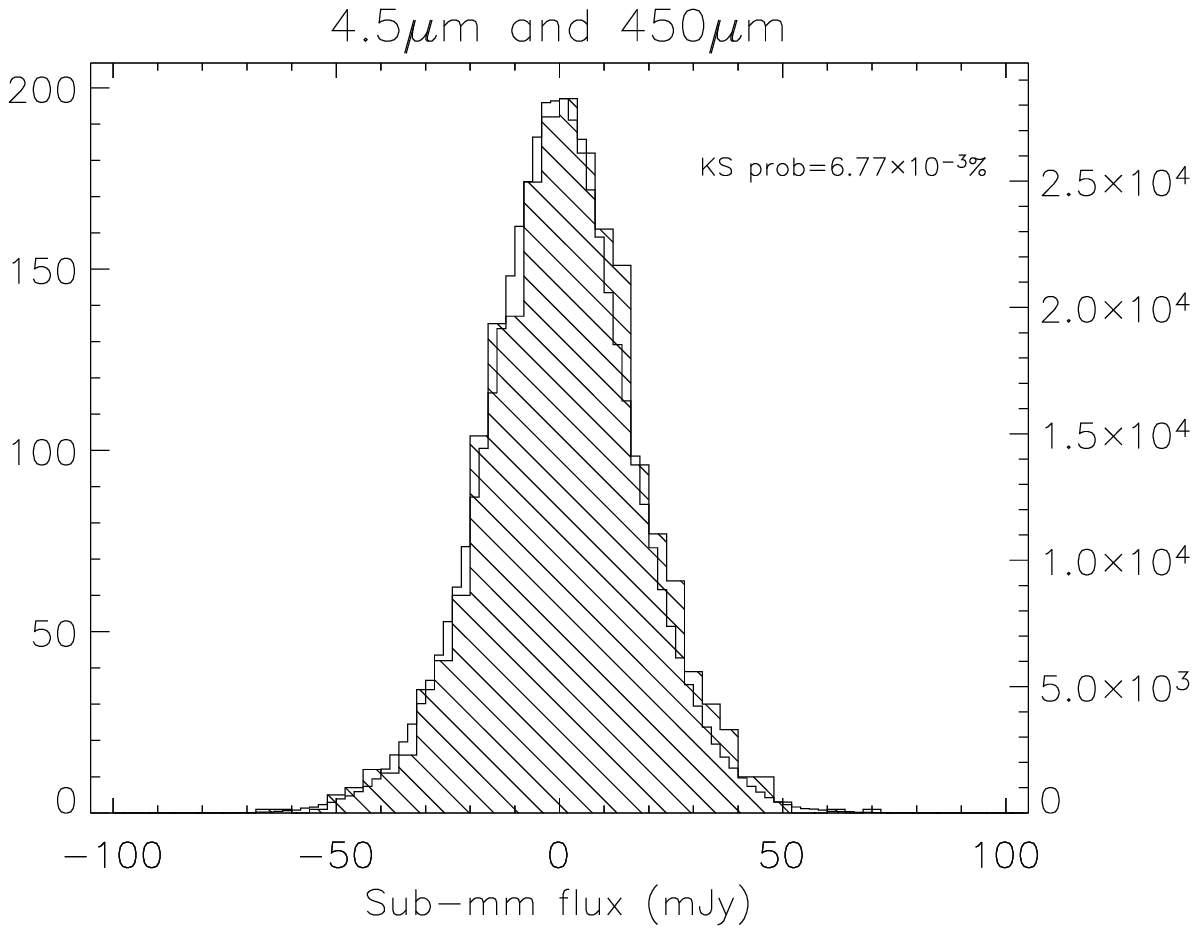}
  \vspace*{-2.5in}\hSlide{2.5in}\ForceHeight{2.5in}\BoxedEPSF{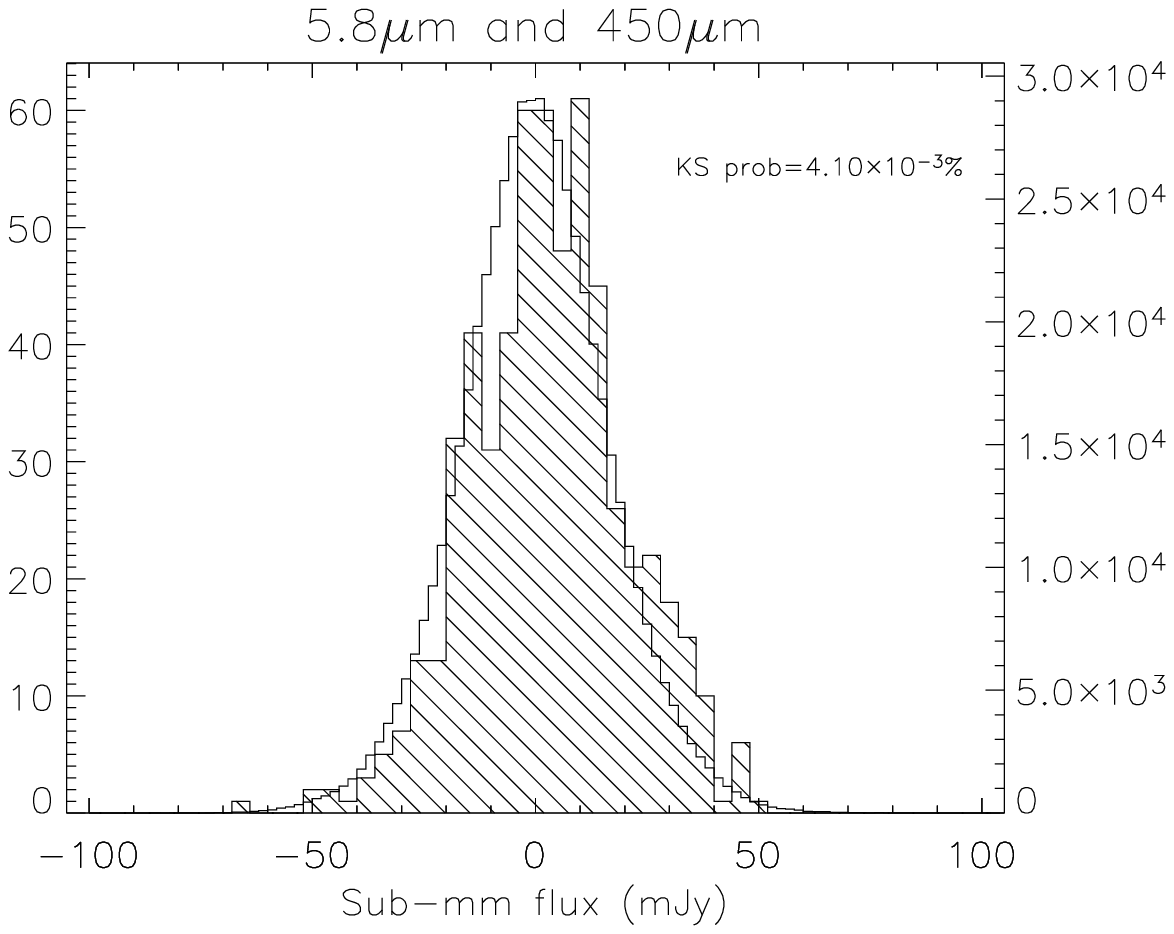}
  \ForceHeight{2.5in}\hSlide{-2.5in}\BoxedEPSF{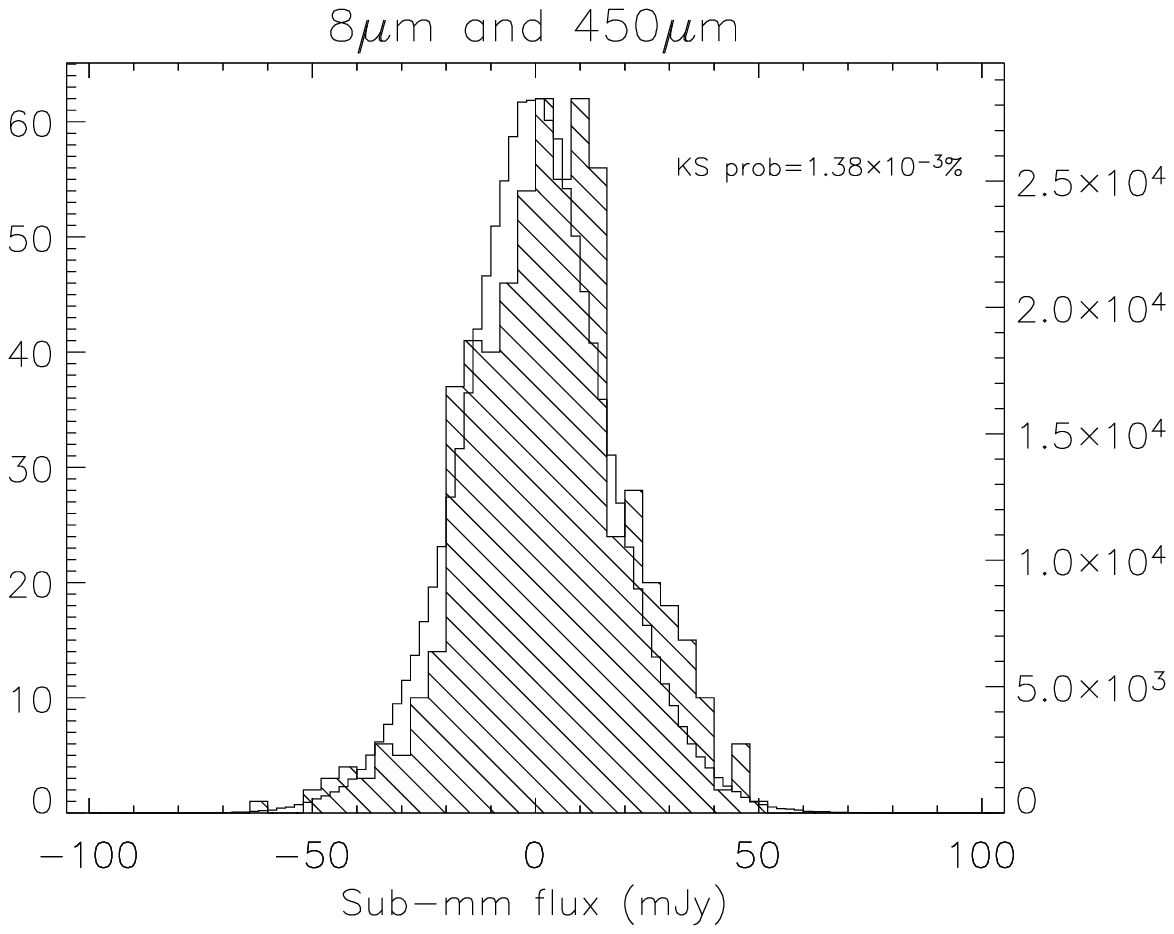}
  \vspace*{-2.5in}\ForceHeight{2.5in}\BoxedEPSF{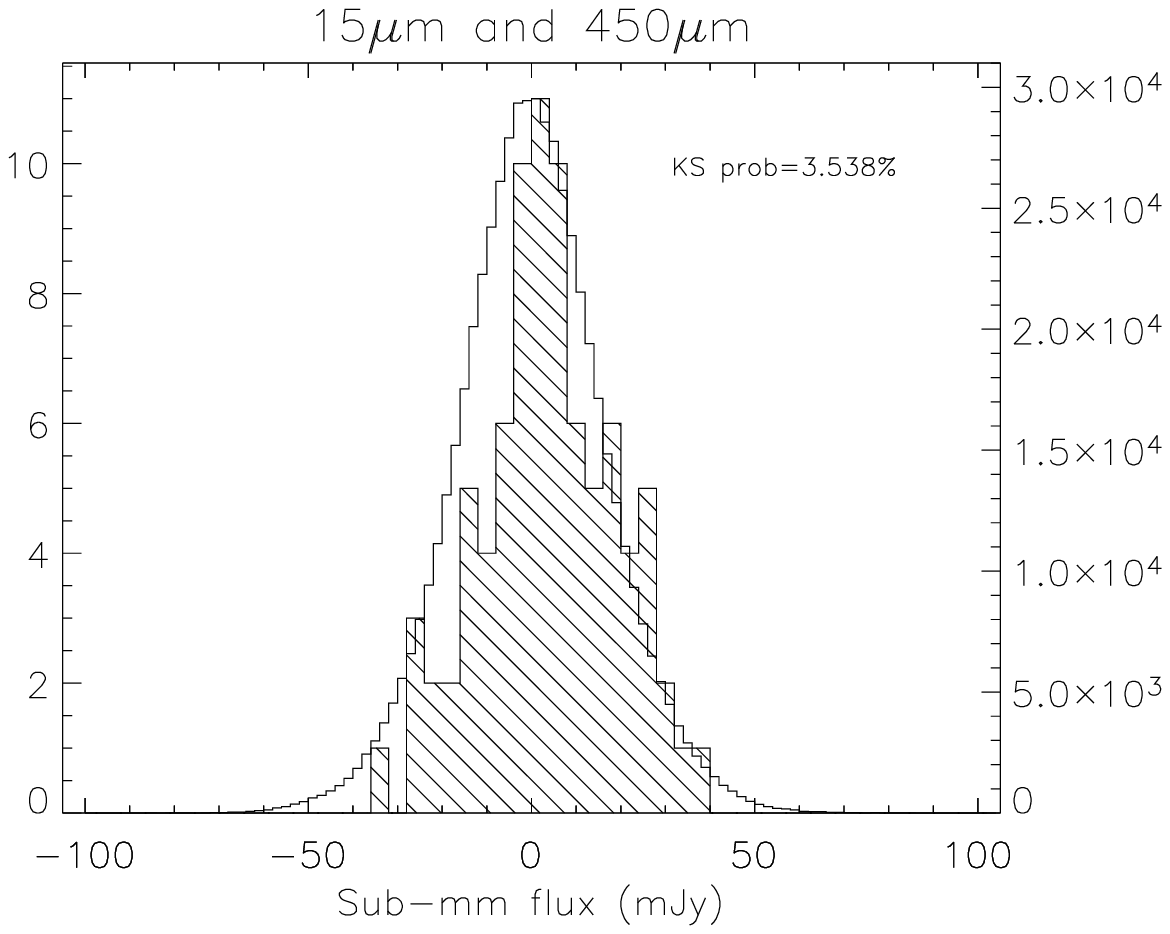}
  \vspace*{-2.5in}\hSlide{2.5in}\ForceHeight{2.5in}\BoxedEPSF{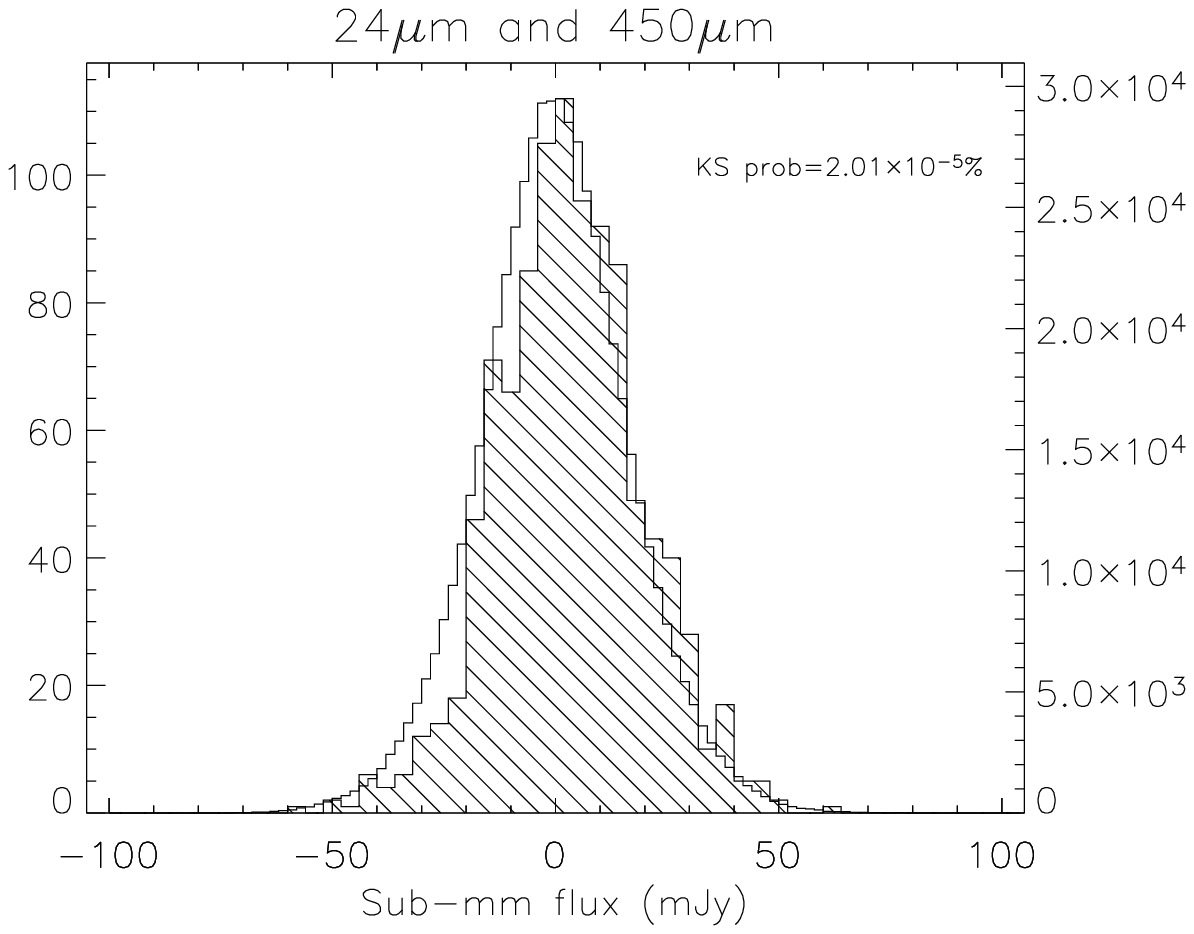}
\caption{\label{fig:450stacks}Histograms of $450\,\mu$m flux in the
  SHADES Lockman map as a whole (unhatched bars) compared to the
  $450\,\mu$m fluxes at the positions of the Spitzer/ISO galaxies selected
  at the indidcated wavelengths (hatched bars). The left-hand vertical scales 
  show the numbers for the hatched histograms, and the right-hand scales refer 
  to the unhatched histograms. Note that in all
  plots, the hatched bars lie typically to the right of the unhatched
  bars. The results of a comparison using the Kolmogorov-Smirnoff test
  are also shown in the diagrams.}
\end{figure*}

\begin{figure*}
\vspace*{-4cm}
\vSlide{-1cm}\ForceWidth{4.5in}\BoxedEPSF{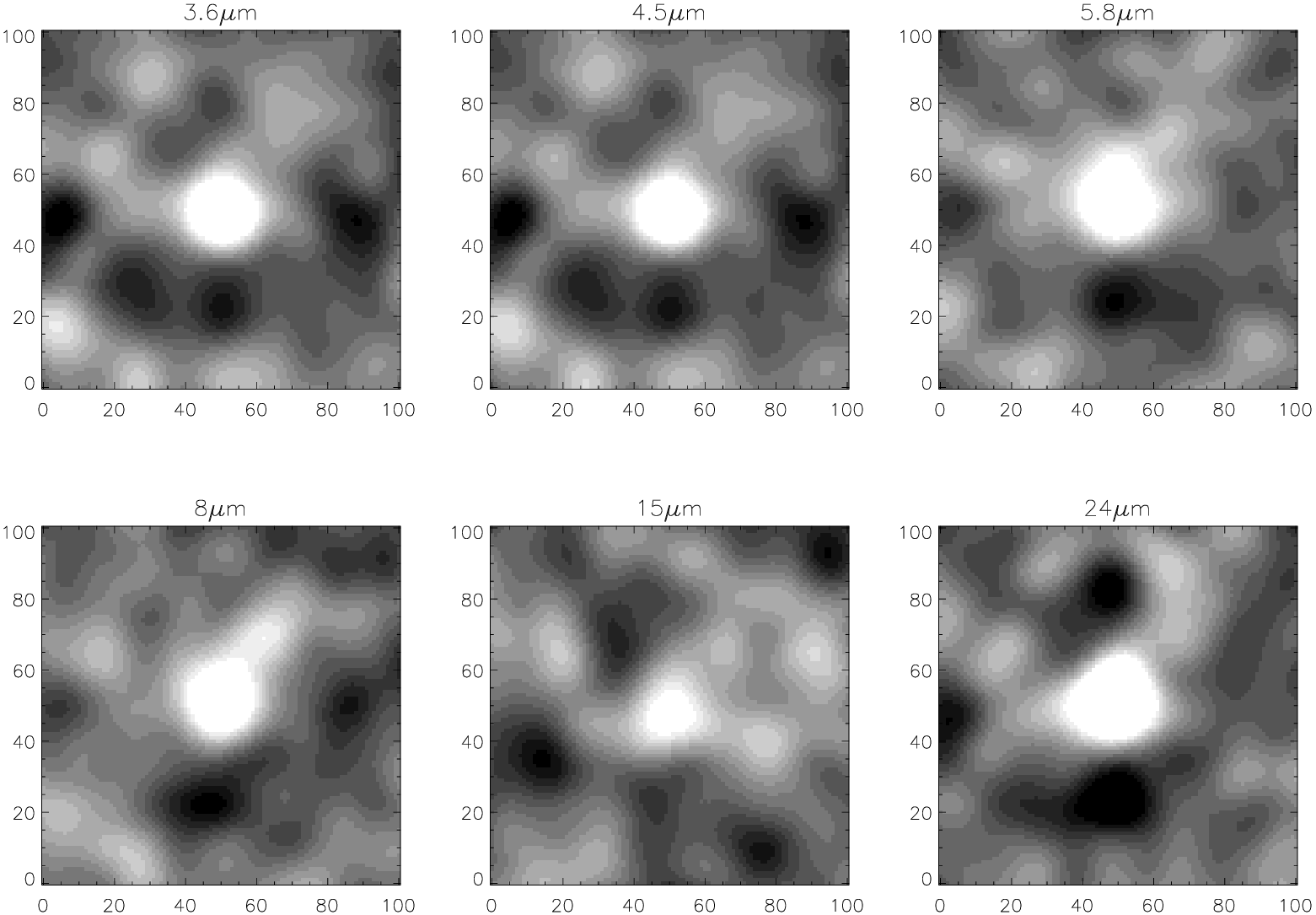}
\caption{\label{fig:weighted_stacks_850}Signal-to-noise images of
  stacked $850\,\mu$m postage stamps centred on Spitzer/ISO galaxies, as
  discussed in the text. The greyscale is from $-3\sigma$ to
  $+4\sigma$. The scales marked are arcseconds. Note the clear
  detections at the centres of all the images. 
  Negative sidelobes are also visible.}
\end{figure*}

\begin{figure*}
\vspace*{-4cm}
\vSlide{-1cm}\ForceWidth{4.5in}\BoxedEPSF{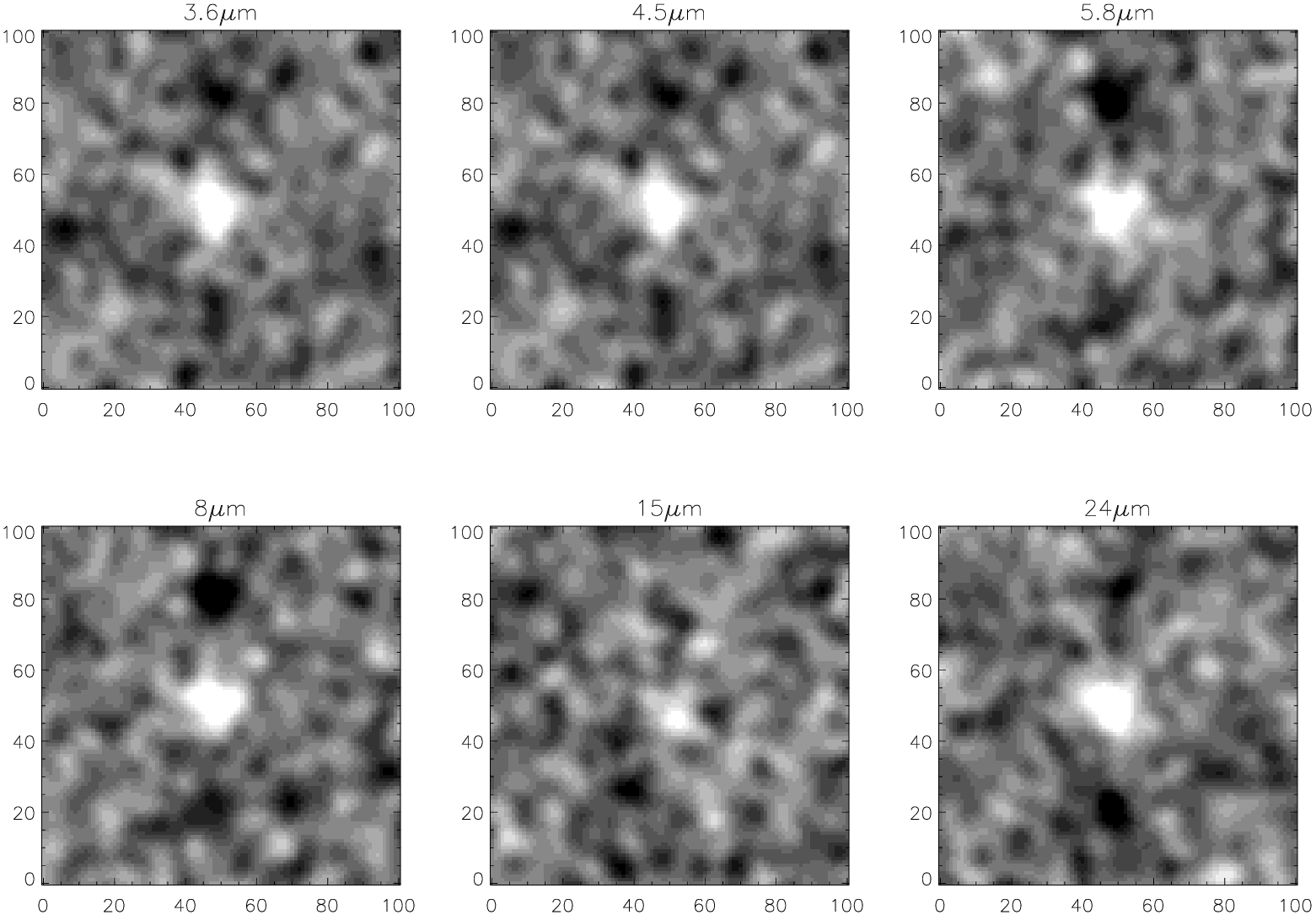}
\caption{\label{fig:weighted_stacks_450}Signal-to-noise images of
  stacked $450\,\mu$m postage stamps centred on Spitzer/ISO galaxies, as
  discussed in the text. The greyscale is from $-3\sigma$ to
  $+4\sigma$. The scales marked are arcseconds. Note the clear
  detections at the centres of all the images. 
  Negative sidelobes are also visible.}
\end{figure*}

\section{Data acquisition}\label{sec:data}
The submm data were taken from 2003-2005 at the James Clerk Maxwell
Telescope (JCMT) with the SCUBA camera, in submm opacities of
$0.265<\tau_{850\mu{\rm m}}<0.283$ and $1.41<\tau_{450\mu{\rm
    m}}<1.52$, i.e. JCMT weather bands 2-3. Chopping/nodding was
performed at position angles of $0$ and $90$ degrees, with chop throws
of $30''$, $44''$ and $68''$, though at the centre of the map only the
$30''$/$90$ degree combination was used (Scott et al. 2002). 
The submm data acquisition,
calibration, reduction and analysis are described in full in Mortier
et al. (2006). A noise-weighted point source filtering was made on the
maps, and the separate chop/nod images were combined optimally
(Serjeant et al. 2003a, Mortier et al. 2006).  
Note that the
SCUBA $450\,\mu$m absolute flux calibration is typically
uncertain to $\sim30\%$. 
The maps from this analysis are used in this paper. Three further
parallel data reduction efforts are described in Coppin et al. (2006),
and $850\,\mu$m maps from some of these reductions have been used to
test the robustness of the results presented in this paper. The
$450\,\mu$m maps used in this paper are from the SHADES data reduction
described as the ``primary'' reduction in Mortier et al. (2005), and
analysis ``B'' in the later Coppin et al. (2006). These maps have the
lowest noise of the available $450\,\mu$m maps, and have short
timescale opacity variations modelled using the water vapour meter.
No reliable $450\,\mu$m point sources are detected in SHADES, though
this is not to say that one cannot find point source candidates in the
maps. A cross-comparison of the four reductions found very few
overlaps between the candidate source lists; this demonstrated that
the $450\,\mu$m opacity during the SHADES runs (see appendix A of
Coppin et al. 2006), and the stability of
the opacity, were not suited to reliable point source extraction. We
will show that this does not preclude statistical constraints on the
$450\,\mu$m-emitting populations.  The $850\,\mu$m maps of the four
data reduction methods are in excellent agreement (Coppin et
al. 2006), and the SHADES point source list is derived from the
consensus of the four analyses.

The Spitzer data were taken in Guaranteed Time, using the IRAC and
MIPS instruments (Fazio et al. 2004, Rieke et al. 2004). As shown in
figure \ref{fig:coverage}, most of the SHADES survey area in the
Lockman Hole East is covered at $3.6\,\mu$m ($4.47\mu$Jy, $3\sigma$),
$4.5\,\mu$m ($4.54\mu$Jy, $3\sigma$), $5.8\,\mu$m ($20.9\mu$Jy,
$3\sigma$), $8\,\mu$m ($12.5\mu$Jy, $3.2\sigma$) and $24\,\mu$m
($38\mu$Jy, $4\sigma$).  We only use Spitzer IRAC sources detected in
at least two Spitzer bands. The area was also mapped at $70\,\mu$m
and $160\,\mu$m, and the comparison between this data and SHADES will
be the subject of a future paper (Egami et al. in preparation). Most
of the SHADES Lockman field was also covered by a deep $15\,\mu$m
survey with the CAM instrument on the Infrared Space Observatory,
further details of which can be found in Elbaz et al. 1999 and
Rodighiero et al. 2004. We select galaxies with $15\,\mu$m flux
densities above $100\mu$Jy for this paper. The Subaru-XMM Deep Field
was also observed by SHADES, and has Spitzer data from the SWIRE
survey (Lonsdale et al. 2004), but since this data is significantly
shallower than the Lockman data we do not consider it here. 

$BRIz$ imaging was obtained from the SUPRIMECAM instrument on
the Subaru telescope, to $5\sigma$ point source depths of
26.8, 25.8, 25.7 and 25.0 in $B$, $R$, $I$ and $z$ respectively
($3''$ diameter AB magnitudes). K-band photometry was
obtained from the UKIRT Deep Infrared Sky Survey (UKIDSS, Lawrence et
al. 2007) to a point source sensitivity of $22.9$ ($5\sigma$ AB
magnitude). Further details are in Dye et al. (2008, paper VII). 

\section{Methodolody and results}\label{sec:method}
\subsection{Photometric redshift estimates}\label{sec:method:photoz}
We use the Spitzer galaxy photometric redshift catalogue of Dye et
al. (2008). This catalogue is derived using the HYPER-Z code
(Bolzonella et al. 2000) applied to the nine-band optical-IRAC
photometric catalogue. Further details of the spectral energy
distribution templates are found in Dye et al. (2008). In figure
\ref{fig:mstar_vs_z} we show the model stellar masses (derived below)
as a function of the photometric redshifts. Note that redshift
aliasing can scatter galaxies to erroneously high redshifts (and hence
to high masses), and there are plausible examples of this in figure
\ref{fig:mstar_vs_z}. We have opted not to impose any arbitrary cuts
in the photometric redshift catalogue to remove these outliers, and
instead leave this to the discretion of the reader. Redshift aliasing 
and consequent erroneously high stellar masses in a small subset would 
not alter the statistical conclusions of this paper. 

Dye et al. (2008) also uses optical and Spitzer photometery to derive
photometric redshift estimates for the SHADES galaxies themselves, and
comparisons with other redshift estimators can be found in Dye et
al. (2008), Aretxaga et al. (2007) and Clements et al. (2007). We will
use the Dye et al. (2008) determinations in this paper. The main
disadvantage of the photometric redshifts in the Lockman Hole is the
lack of spectroscopic training sets, though a comparison of the Dye et
al. photometric redshifts of submm-selected galaxies with their
spectroscopic redshifts, and between independent photometric redshift
determinations (their figures 3 and 4) shows the photometric
redshifts are accurate to $|\Delta z|/(1+z)\simeq0.09$ consistent
with other studies (e.g. Chapman et al. 2005, Pope et al. 2006). 
This is more than sufficient for our purposes. 

\subsection{Mass estimates of Spitzer galaxies}\label{sec:method:mass}
The $3.6\,\mu$m and $4.5\,\mu$m IRAC bands are dominated by the redshifted
light from old stellar populations, and are therefore useful
estimators of the assembled stellar masses. The K-corrections are also
simple at these wavelengths, since it samples the Rayleigh-Jeans tail
of the stellar photospheric emission. We use the model spectral energy
distributions from Dye et al. (2008) to obtain
rest-frame K-band monochromatic luminosities; the results are
insensitive to the assumed spectral energy distribution. 

There is no accepted conversion between rest-frame K-band luminosity
and stellar mass, as a function of redshift. Our aim is to compare our
results with the de Lucia (2005) Millenium simulation, and one can
only do this self-consistently by adopting a conversion consistent
with that simulation. Therefore, we adopt an empirical conversion
based on these simulations, based on fits to the simulated data in
figure \ref{fig:millenium_conversion}.  This conversion is consistent
with the observed luminosity-dependent evolution in K-band stellar
mass-to-light ratios in the MUNICS survey (Drory et al. 2004). Our
conversion is 
\begin{equation}\label{eqn:kmconv}
\log_{10}M_*=a(z)+b(z)\times K_{\rm abs, AB}
\end{equation}
 where the values of $a$
and $b$ are interpolated from the best fit values tablulated in table
\ref{tab:kfit}. 

In calculating the matter overdensities, we add a dark matter
contribution following the total mass estimates from gravitational
lenses by Ferreras et al. 2005, who found that the total mass scales
as the stellar mass to the $1.2$ power, with total mass equalling
stellar mass at $3.18\times10^9 M_\odot$. Below this mass we make no
correction for dark matter contribution. Our results are not sensitive
to the dark matter assumptions. 

\subsection{The density field around submm galaxies}\label{sec:method:density}
As shown in Blake et al. (2006), the SHADES survey does not have
enough sources or field galaxy redshifts to accurately determine the
galaxy-SHADES cross-correlation function. We therefore 
used an alternative
estimator 
which is essentially a stack of the $3.6\,\mu$m Spitzer data at the positions 
of the SHADES sources. 
Using Spitzer galaxy mass estimates discussed above, 
we created projected mass density maps in broad redshift
bins, and smoothed each map with a top-hat circular kernel with
$0.5-2$ comoving Mpc diameters 
at the central redshift of the bin. We
then compared the projected matter density at the positions of the
SHADES galaxies with the histogram of matter density for the map as a
whole, using methods similar to established techniques for stacking
analyses (see e.g. section \ref{sec:method:stacking}). We omitted
SHADES galaxies lying in regions with zero density 
(no $3.6\,\mu$m identification nor sufficiently close neighbours) and
also restricted the comparison to the non-zero density regions of the
map as a whole. 

This comparison 
is shown 
in figures \ref{fig:percentiles} and \ref{fig:percentiles_2Mpc} where
the percentile of the submm galaxies' environment is plotted against
redshift. In the $0.5\,$Mpc smoothing case, the submm galaxies in the
lower redshift bin lie in the top $10-20$ percentile of the matter
density distribution. However, the small smoothing kernel leaves large
regions of the map with no density field data, so the sample of submm
galaxies is small. In the $2\,$Mpc smoothing case, the sample size is
more than doubled; only five galaxies are excluded in the lowest
redshift bin.  The submm galaxies appear to lie in a wide range of
environments, but around a third of the submm galaxies at $1<z<1.5$
lie in the top $15$ percentile of the density distribution. 

We performed a similar calculation for $3.6\,\mu$m-selected galaxies,
which are plotted as small dots in figures \ref{fig:percentiles} and
\ref{fig:percentiles_2Mpc}. These appear to have a more uniform
distribution in the lower redshift bin than the submm galaxies. We
compared the submm galaxy and $3.6\,\mu$m-selected populations in the
$1<z<1.5$ bin using a Kolmogorov-Smirnoff test. In the $2\,$Mpc
smoothing case the difference is only marginally significant ($32\%$
probability that the distributions are the same), but in the
$0.5\,$Mpc case the distributions are clearly different ($0.2\%$
probability that the distributions are the same). Note that in making
this comparison, we have subtly changed the question we are asking. We
wish to know if submm galaxies lie in richer environments than
average, but this average can be taken as a volume average or as a 
per-galaxy average. The raw percentiles address the former, and the
submm-Spitzer comparision addresses the latter. 

\subsection{The submm emission of $3-24\,\mu$m Spitzer
  galaxies}\label{sec:method:stacking} 
In Serjeant et al. (2004), the regions of the map near submm point
sources were simply excluded from the analysis. This runs the risk of
removing the submm signal from companions to the submm point
source. Here, point sources detected in the combined
point-source-filtered $850\,\mu$m map with significance levels of
$\ge3.5\sigma$ were subtracted from the original chop/nod images,
which were then filtered with the chopped/nodded point source kernels,
and the maps were optimally combined to create a residual $850\,\mu$m
image. This is the same as the procedure adopted by Dye et al. (2006), 
and this is the threshold used for the submm point source catalogue in 
Coppin et al. (2006). (We will show below in figures
\ref{fig:850stacks} and \ref{fig:450stacks} that our results are not
sensitive to this threshold.) 
There is evidence that the submm point source population has
different Spitzer:submm flux ratios than the Spitzer-selected
population (Serjeant et al. 2004), so it is important to remove the
point source population before stacking.  No reliable $450\,\mu$m point
sources are detected, so the point-source-filtered maps at $450\,\mu$m
are used without modification. Our methodology differs from that of 
Wang et al. (2006), in which submm point sources were left in the
map.

The submm point spread function sums to exactly zero, because of the
negative sidelobes from the chopping and nodding. Therefore, there is
no risk of overestimating the submm flux of a given Spitzer galaxy by
also counting its neighbours (Peacock et al. 2000, Serjeant et
al. 2004). This is because the expectation value of the submm flux
from (unclustered) neighbouring galaxies equals that of the map, which
is exactly zero by virtue of the zero-sum point spread function. The
effect of the clustering of the Spitzer population is estimated below
in section \ref{sec:method:spitzerclustering}.

We can therefore calculate the mean submm flux of Spitzer galaxies by
averaging the measurements at the positions of Spitzer galaxies in the
submm maps, even if there is $>1$ Spitzer galaxy per submm beam. The
difficulty in extracting $450\,\mu$m point sources (section
\ref{sec:data}) raises the possibility of non-Gaussian features in the
maps contributing to the signal. Such non-Gaussian features could be
caused by, for example, imperfect sky subtraction or imperfect
correction for atmospheric opacity; the sky is many orders of
magnitude brighter than the extragalactic signal (e.g. Serjeant et
al. 2003a). However, by the Central Limit Theorem, the probability
distribution of the {\it mean} of a sample is approximately Gaussian
with a variance $\sigma^2/n$ where $\sigma^2$ is the variance of the
underlying distribution being sampled; the distribution of the mean is
increasingly Gaussian for larger samples and for more Gaussian-like
underlying distributions. Here, the underlying distribution is
well-approximated as Gaussians (e.g. Mortier et al. 2005), and many
thousands of samplings are taken from the underlying distribution. We
are therefore confident that the mean flux levels of stacked
populations are Gaussian-distributed.

The mean flux level has the advantage of being physically
interpretable, but it is not 
necessarily 
the most efficient detection statistic of a stacked
signal. The Kolmogorov-Smirnoff test has been widely used to test
whether the distribution of submm fluxes at the positions of interest
are a random sampling from the map as a whole (e.g. Serjeant et
al. 2004, Dye et al. 2006).  This test is asymptotically
distribution-free, and is therefore insensitive to non-Gaussian
features in the underlying maps. Furthermore, a comparison with a control
sample (such as randomized submm source positions) is intrinsic to the
test, since it compares the flux distribution of the map as a whole
with the map fluxes at the positions of interest.  It is also possible
to translate the Kolmogorov-Smirnoff significance level into an
equivalent number of $\sigma$ of a Gaussian distribution, by inverting
$P_{\rm KS}=0.5{\rm erf}(\sigma/\sqrt{2}) + 0.5$.

Some authors have advocated the use of the error-weighted mean fluxes,
rather than the unweighted mean stacked fluxes (e.g. Dye et
al. 2006). In Serjeant et al. (2004) it was argued that it was not
obvious that the Central Limit Theorem would apply to these weighted
quantities. However, Dye et al. (2006) tested their weighted means and
found them to be only subtly biased. The large area of the SHADES maps
may provide an advantage: provided the field of view is sufficiently
large, and provided the noise level is sufficiently uniform, the
zero-sum point spread function of each individual source will still
produce a zero net contribution on average to the $S/N$
(signal-to-noise) and $S/N^2$ images, as well as to the flux image
$S$. In this paper we will make error-weighted coadded submm postage
stamps of the sources to be stacked. We will show that our
error-weighted mean fluxes show no evidence for a systematic shift
relative to the unweighted mean fluxes.

We applied the methodology of Serjeant et al. (2004) to test whether
the submm fluxes at the positions of Spitzer/ISO galaxies were
representative of the submm map as a whole, or whether there is on
average positive flux at the Spitzer/ISO galaxy positions. We excluded
regions of the $850\,\mu$m map with noise levels above $5$mJy, and
regions of the $450\,\mu$m map with noise levels above $20$mJy, since
these disproportionately affect the non-noise-weighted stacks. At both
wavelengths, we then calculated the median noise levels in the
unmasked regions, and then masked all areas with more than twice these
noise levels. The $450\,\mu$m data quality is much more dependent on
the weather conditions, so more of the $450\,\mu$m map is excluded by
our noise cuts (e.g. table \ref{tab:meanfluxes}).  The results are
shown in figures \ref{fig:850stacks} and \ref{fig:450stacks}. We
detect the Spitzer galaxies at $\gg99\%$ confidence at both
$450\,\mu$m and $850\,\mu$m, and for all Spitzer wavelengths from
$3-24\,\mu$m. We also detect the ISO $15\,\mu$m-selected population at
$99.6\%$ confidence at $850\,\mu$m and $96\%$ confidence at
$450\,\mu$m. The mean fluxes are given in table \ref{tab:meanfluxes},
as are the Gaussian-equvalent $\sigma$ values of the
Kolmogorov-Smirnoff significance levels (though recall the
uncertainties in the $450\,\mu$m flux calibration noted above).  Our
stacked mean fluxes are in good agreement with the previous
determinations of Serjeant et al. (2004), though in addition we have
made clear submm stacking detections at $3.6\,\mu$m and
$4.5\,\mu$m. The mean flux ratios for the $5.8-8\,\mu$m-selected
galaxies presented in table \ref{tab:meanfluxes} are somewhat lower
than those presented in Serjeant et al. (2004), which may be due to
the lack of brighter Spitzer sources in the very small field of view
of the Early Release Observations. There are some difficulties in
interpreting these flux ratios, as they are summed from galaxies
spanning a range of redshifts, and the contribution made by individual
galaxies will depend on their location in the luminosity-redshift
plane. Furthermore, the mean mid-infrared flux is sensitive to the
presence of the few brighter sources in the sample, which may lead to
underestimates in the quoted errors in the flux ratios. We will return
to this topic in section \ref{sec:discussion}. Another anomalous flux
ratio is the $450\,\mu$m:$15\,\mu$m ratio; we believe this is due
partly to small number statistics in this sample, and partly to the
fact that only brighter mid-infrared galaxies are detected at this
wavelength which may bias the sample to submm-weak AGN dust
tori. These galaxes are also only marginally detected (table
\ref{tab:meanfluxes}).

\begin{figure}
  \ForceWidth{4.5in}
  \hSlide{-1cm}
  \BoxedEPSF{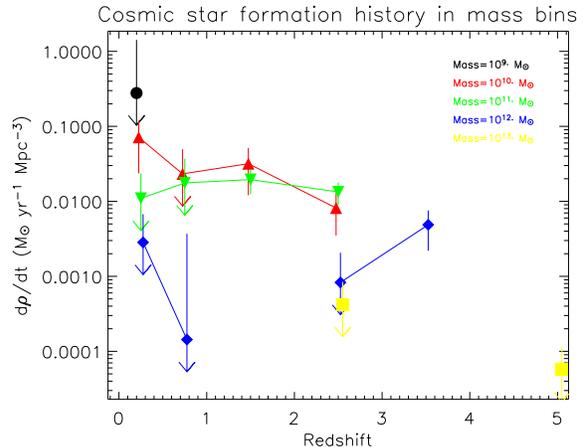}
\caption{\label{fig:madau} Evolution in comoving volume-averaged star
  formation rates as a function of total mass for $3.6\,\mu$m-selected
  galaxies, in $\pm0.5$\,dex mass bins centred on the masses
  shown. The brighter submm point source population
  discovered by Smail et al. 1997, Hughes et al. 1998 and Barger 
  et al. 1998 are not
  included, because we removed submm point sources prior to our
  stacking analysis. Symbols: the circle is $10^9 M_\odot$, upward triangles
  $10^{10} M_\odot$, downward triangles $10^{11} M_\odot$, diamonds
  $10^{12} M_\odot$, and the square is the $10^{13} M_\odot$
  constriant. Note that the lower-mass galaxies have star formation
  histories peaking at much lower redshifts than the putative
  $z\simeq2.2$ peak in the bright submm point source population.}
\end{figure}

The Kolmogorov-Smirnoff test reports an apparently unrealistically
small significance when comparing the $850\,\mu$m fluxes of
$24\,\mu$m-selected galaxies with the submm map as a whole. Examination
of the numbers in the bins of the histograms shows why this is the
case; in effect, the probability that the distributions are identical
is immeasurably small.  To our knowledge, this is the best submm
stacking detection ever made.

We constructed a noise weighted sum of the postage stamps around
Spitzer/ISO galaxies selected at each wavelength.  Figures
\ref{fig:weighted_stacks_850} and \ref{fig:weighted_stacks_450} show
the signal-to-noise images of the galaxies selected at these
wavelengths. We clearly have strong detections of our sample at all
wavelengths. Furthermore, the off-centre positions in these stacked
postage stamps provide a control, and confirm the stacking is not
prone to false positives. The weighted means are quoted in table
\ref{tab:meanfluxes} and are in good agreement with the unweighted
means.  In particular, there is no evidence in table
\ref{tab:meanfluxes} for weighted means systematically offset from the
unweighted mean fluxes; any systematic offset must be far smaller than
the random noise in the measurements.

\begin{table*}
\begin{tabular}{lllllll}
                         & $3.6\,\mu$m & $4.5\,\mu$m & $5.8\,\mu$m & $8\,\mu$m & $15\,\mu$m & $24\,\mu$m\\
\hline 
$N_{850}$                 & $4803$ & $4676$ & $1505$ & $1558$ & $188$ & $2656$\\
$\langle S_{850}\rangle$/mJy  & $0.184\pm0.029$ & $0.181\pm0.029$ & $0.302\pm0.052$ & $0.300\pm0.051$ & $0.43\pm0.14$ & $0.361\pm0.041$\\
$\langle S_{850}\rangle_{\rm weighted}$/mJy &  
                         $0.195\pm0.024$ & $0.191\pm0.024$ & $0.318\pm0.043$ & $0.292\pm0.042$ & $0.54\pm0.12$ & $0.343\pm  0.033$\\
$\langle S_{850}\rangle/\langle S_{\rm mid-IR}\rangle$ & 
                         $2.64\pm0.24$ & $3.49\pm0.37$ & $1.92\pm0.27$ & $2.80\pm0.36$ & $0.74\pm0.08$ & $1.98\pm0.06$\\
$\sigma_{{\rm KS}, 850}$   & $5.3\sigma$  & $5.9\sigma$ & $4.6\sigma$ & $5.3\sigma$  & $2.6\sigma$ & $\gg7\sigma$\\
$N_{450}$                 & $1994$ & $1977$ & $586$ & $625$ & $84$ & $1023$\\
$\langle S_{450}\rangle$/mJy  & $1.23\pm0.38$ & $1.23\pm0.39$ & $2.98\pm0.72$ & $2.84\pm0.70$ & $4.1\pm1.6$ & $3.00\pm0.51$\\
$\langle S_{450}\rangle_{\rm weighted}$/mJy &  
                          $1.55\pm0.27$ & $1.56\pm0.27$ & $3.00\pm0.49$ & $2.95\pm0.47$ & $3.6\pm1.3$ & $2.53\pm0.36$\\
$\langle S_{450}\rangle/\langle S_{\rm mid-IR}\rangle$ & 
                         $45.2\pm5.1$ & $72.6\pm7.3$ & $14.7\pm3.8$ & $30.0\pm5.4$ & $1.58\pm0.2$ & $11.5\pm0.4$\\
$\sigma_{{\rm KS}, 450}$   & $3.8\sigma$  & $3.8\sigma$ & $3.9\sigma$ & $4.2\sigma$  & $1.8\sigma$ & $5.1\sigma$\\
\end{tabular}
\caption{\label{tab:meanfluxes} Mean submm fluxes for Spitzer-selected
  and ISO-selected populations. Also quoted are the numbers of
  galaxies in each stacking analysis. Kolmogorov-Smirnoff significance
  values are quoted in figures \ref{fig:850stacks} and
  \ref{fig:450stacks}, and translated in this table into an equivalent
  number of $\sigma$ of a Gaussian distribution as discussed in the
  text, except for the $850\,\mu$m detection of $24\,\mu$m sources which
  has a significance level too high for numerical inversion. The
  quoted errors on the mean fluxes are $\sigma/\sqrt{N}$, where $N$ is
  the number of measurements and $\sigma^2$ the variance of the
  measurements, except in the case of flux ratios where the
  $\sigma/\sqrt{N}$ errors in each quantity have been
  propagated. Known submm sources have been subtracted from the maps
  prior to this analysis, as discussed in the text. The effect of the
  clustering of Spitzer galaxies is discussed in section
  \ref{sec:method:spitzerclustering}.} 
\end{table*}

\begin{figure*}
  \ForceWidth{4.5in}
  \hSlide{-4.5cm}
  \BoxedEPSF{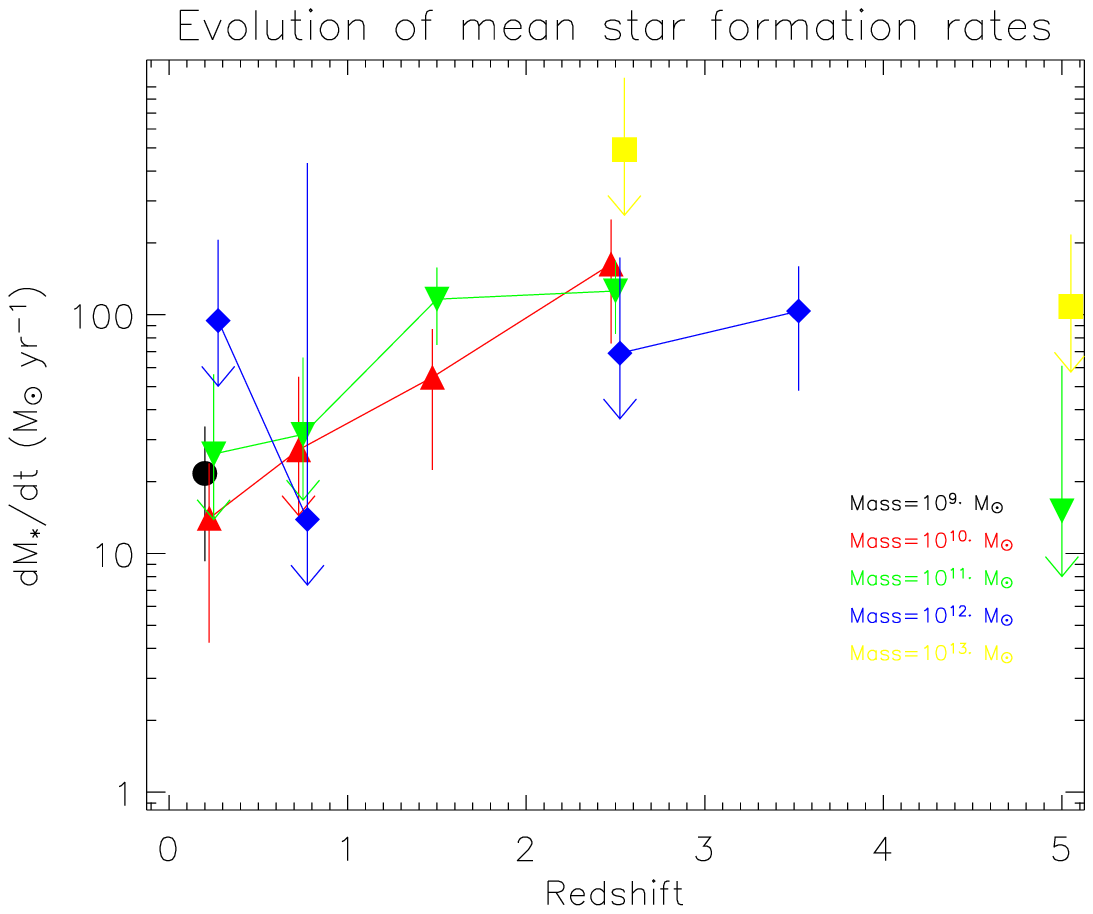}
\vspace*{-8.2cm}
  \ForceWidth{4.5in}
  \hSlide{4.5cm}
  \BoxedEPSF{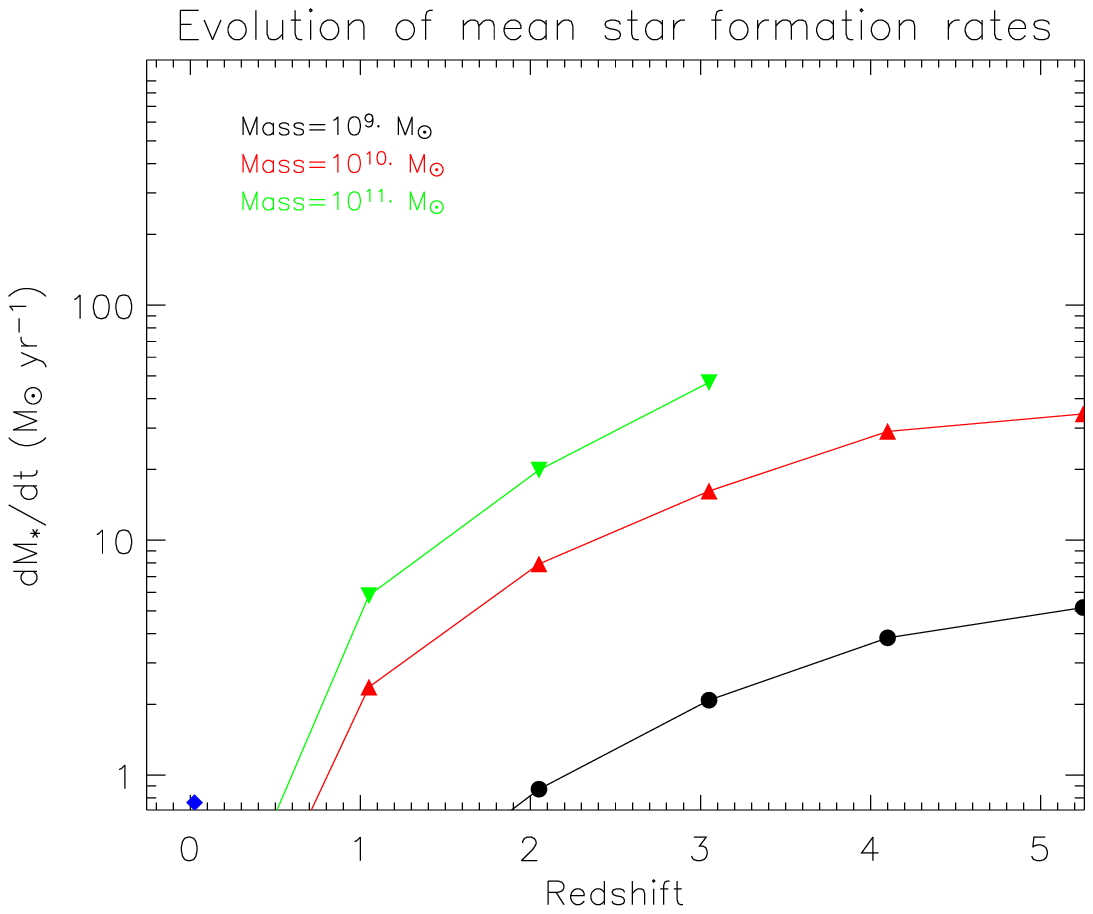}
\caption{\label{fig:mean_sfr} 
  Left: mean star formation rates, $dM/dt$, of our galaxies.
  Symbols as in figure \ref{fig:madau}. Right: corresponding
  predictions from the de Lucia et al. (2006) simulation. Note that
  the observed star formation rates clearly exceed the predictions.}
\end{figure*}

\begin{figure*}
  \ForceWidth{4.5in}
  \hSlide{-4.5cm}
  \BoxedEPSF{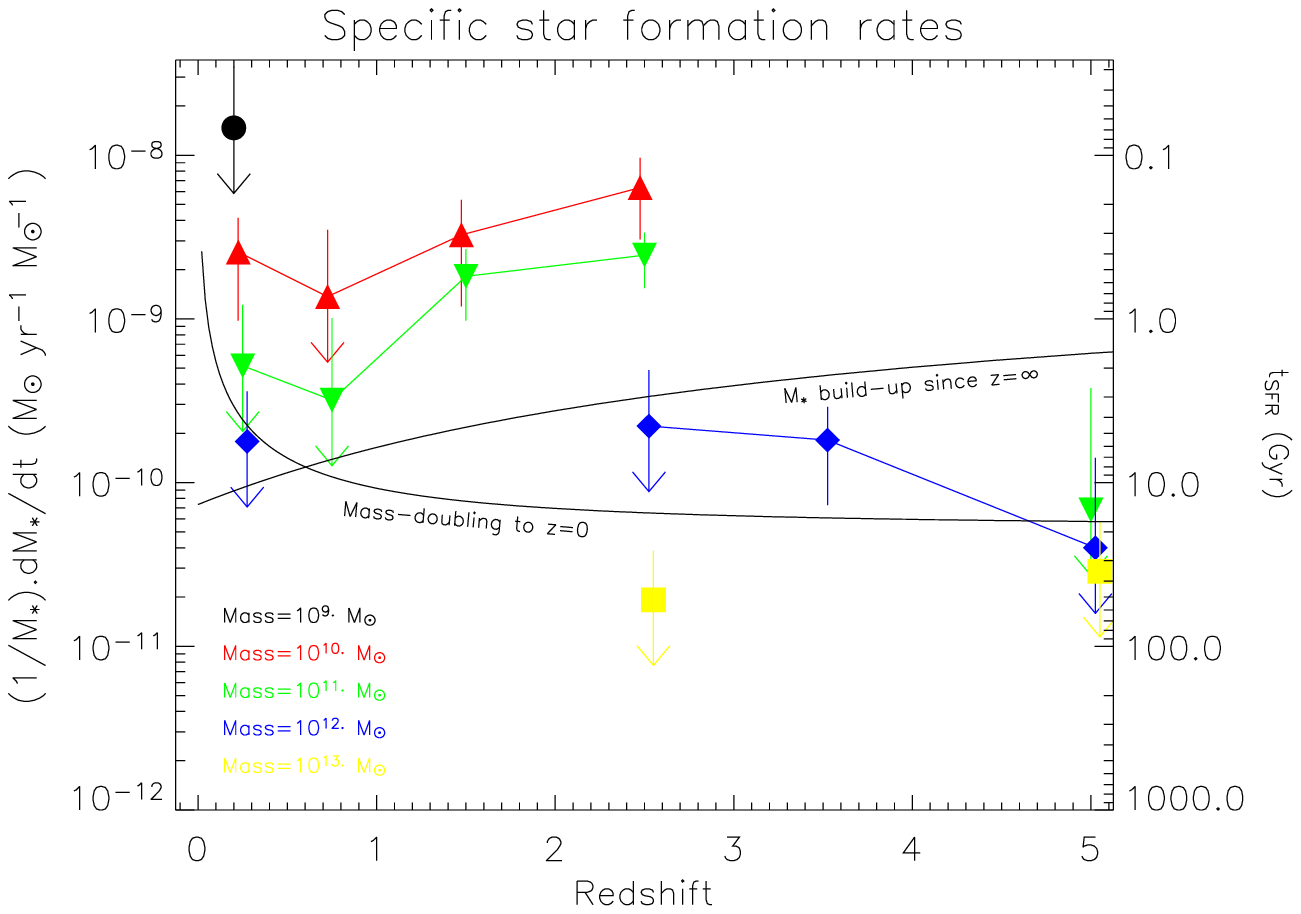}
\vspace*{-8.2cm}
  \ForceWidth{4.5in}
  \hSlide{4.5cm}
  \BoxedEPSF{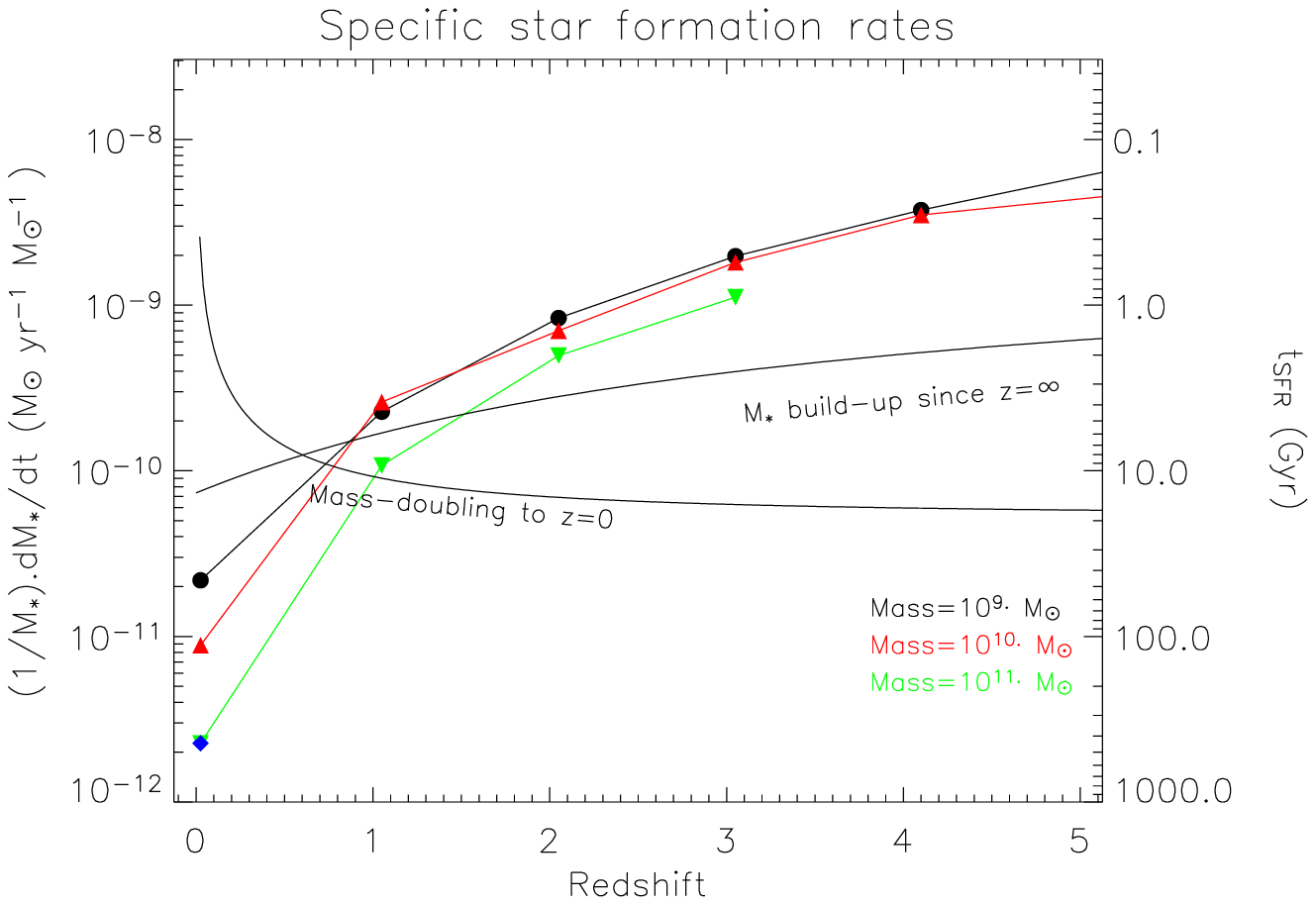}
\caption{\label{fig:specific_sfr} 
  Left: star formation rate per unit galaxy mass, $(1/M)dM/dt$. This has
  dimensions of $[T]^{-1}$ and the corresponding star formation
  timescales are given in the right-hand ordinate. Symbols as in
  figure \ref{fig:madau}. Also plotted is the specific star formation
  rate required to double the stellar mass by $z=0$ assuming a
  constant {\it specific} star formation rate, and the rate required
  to assemble the galaxy since the Big Bang assuming a constant star
  formation rate. Points lying above either line may be regarded as
  starbursting. Note that most of our detections can be regarded as
  starbursting. Right: corresponding predictions from the de Lucia
  et al. (2006) simulation. Note the obvious discrepancies with the
  observations.}
\end{figure*}

\begin{figure}
  \ForceWidth{4.5in}
  \hSlide{-1.5cm}
  \BoxedEPSF{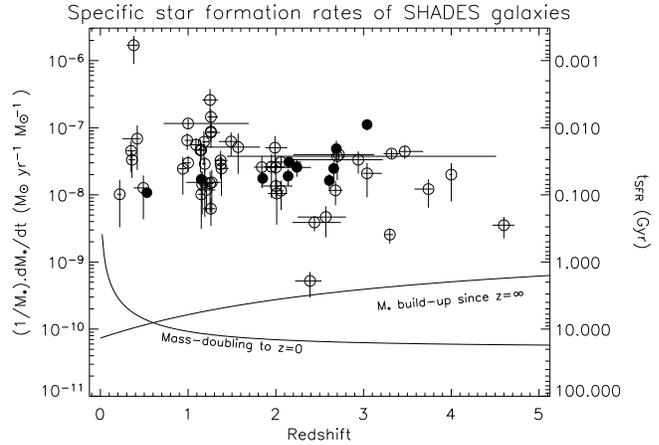}
\caption{\label{fig:shades_specific_sfr} Star formation rate per unit
  galaxy mass, $(1/M)dM/dt$, for the SHADES point sources in the
  Lockman Hole. As with figure \ref{fig:specific_sfr}, the
  corresponding star formation timescales are given in the right-hand
  ordinate.  Objects with spectroscopic redshifts are shown in filled
  symbols, and objects with optical-Spitzer-based photometric
  redshifts from Dye et al. (2008) are shown as open symbols.  Note
  that the star formation timescales for SHADES sources are much
  shorter than in the population as a whole.}
\end{figure}

Since our signal-to-noise is so high (unlike in the much smaller
sample of Serjeant et al. 2004), we can investigate the
sub-populations which dominate the stacking signal. In figure
\ref{fig:madau} we plot the comoving volume-averaged star formation
rate estimated from the $850\,\mu$m stacked fluxes, assuming an M82
spectral energy distribution shape, and using a conversion derived for this
SED from the Kennicutt (1998) conversion:
\begin{equation}
\frac{\rm SFR}{M_\odot/{\rm year}} = 
   \frac{L_{\rm FIR}}{5.8\times10^9L_\odot} = 
   \frac{\nu L_\nu(60\mu{\rm m})}{3.6\times10^9L_\odot}
\end{equation}
This assumes a Salpeter initial mass function from
$0.1\,M_\odot$ to $100\,M_\odot$. Our estimator for the
total submm flux contribution from a population of galaxies in a
logarithmic mass interval $\Delta\log_{10}M$ is
\begin{equation}
F_{\rm tot}(\rm mJy~deg^{-2} dex^{-1})=\frac{1}{A\Delta\log_{10}M}\Sigma_{i=1}^{N} M(i)/c(i)
\end{equation}
where $A$ is the survey area, $M(i)$ is the submm map flux at the
position of galaxy $i$ (of which there are $N$) and $c(i)$ is the
completeness of the Spitzer catalogue for galaxies similar to $i$. The
calculation is not sensitive to the completeness correction. To first
order both the submm and Spitzer fluxes are constant over the redshift
intervals in question (due to the negative K-corrections at both
wavelength ranges). Our estimate of the error on the total flux is
\begin{equation}
\Delta F_{\rm tot}=\frac{1}{A\Delta\log_{10}M}\sqrt{\Sigma_{i=1}^{N} (M(i)/c(i))^2}
\end{equation}
For conversion from $850\,\mu$m fluxes to luminosities, we use the
central redshift of the bin, noting that the submm luminosity is
roughly independent of redshift over the redshift ranges considered.
We do not plot bins in which galaxies with masses equal to the mass at
the centre of the bin are not above the Spitzer flux limit throughout
the bin. We weight the flux contributions of each galaxy 
according to its accessible
comoving volume. Figure \ref{fig:mean_sfr} shows the mean (quiescent) star
formation rate per galaxy. Figure \ref{fig:specific_sfr} shows the
mean (quiescent) star formation rate per unit galaxy mass, as a
function of mass. This quantity has the dimensions of one over time
(e.g. Gyr$^{-1}$). This characteristic star formation timescale can be
regarded as the timescale over which the bulk of the galaxy's baryonic
matter would be converted into stars (though for consistency with
elsewhere in this paper we use the total mass estimates for the
galaxies, not just baryonic).  The quantities plotted in figure
\ref{fig:specific_sfr} have no dependence on the completeness of the
Spitzer catalogue, though the errors on the quantities depend on the
sample size. We have also plotted the mass-doubling timescale as a
function of redshift on this figure, assuming {\it SFR}$/M_*$ is
constant; galaxies above this line may be regarded as starbursting. A
further useful metric, also plotted in this figure, is the specific
star formation required to build up the entire observed stellar mass
since the Big Bang, assuming {\it SFR} is constant. For comparison,
the specific star formation rates of the SHADES point sources are
shown in figure \ref{fig:shades_specific_sfr}.  We will discusss these
figures in section \ref{sec:discussion:mass}.

\subsection{The cosmic near-infrared and submm
  backgrounds}\label{sec:method:background} 

\begin{figure*}
                  \hSlide{-2.0in}\ForceHeight{2.5in}\BoxedEPSF{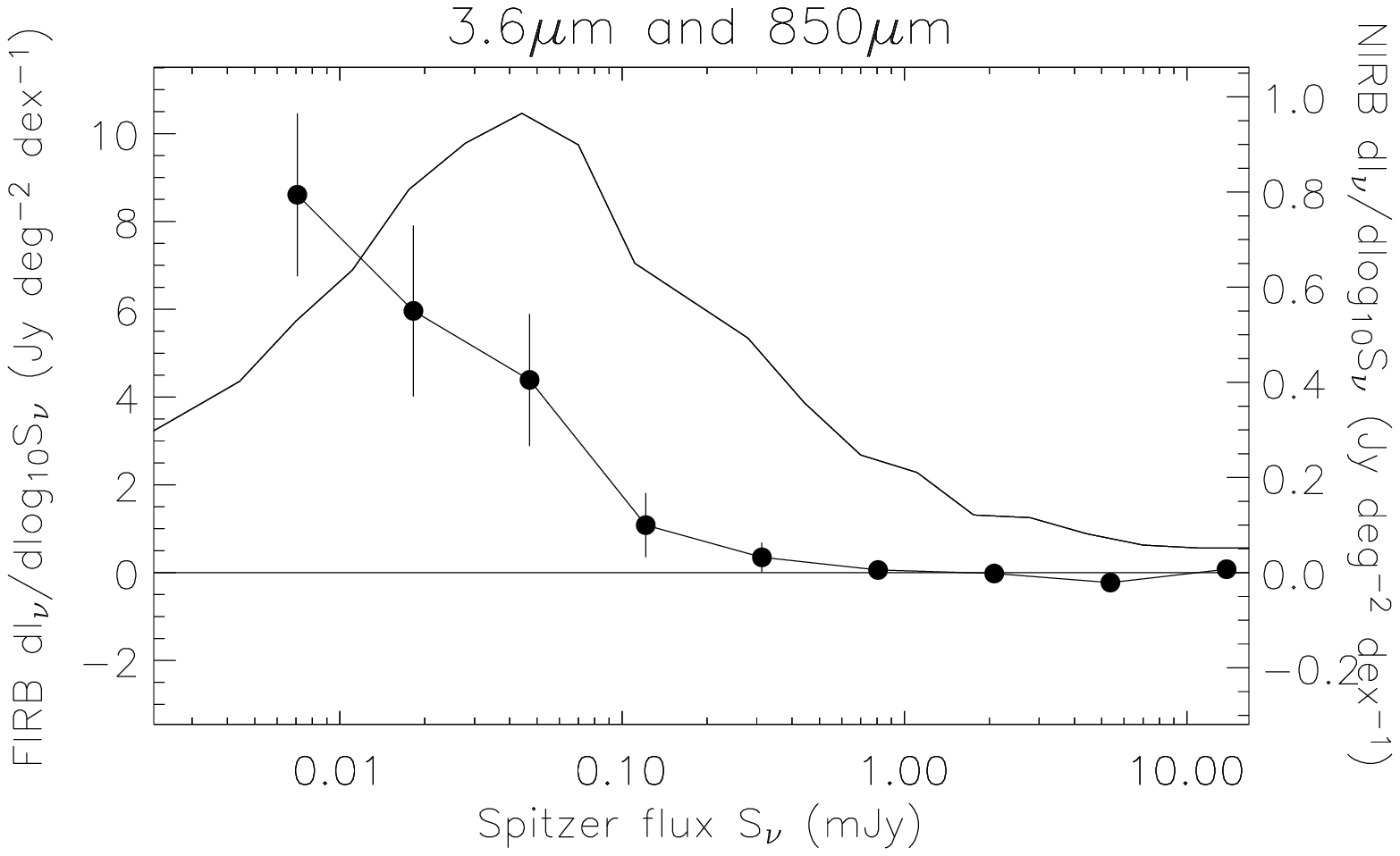}
  \vspace*{-2.5in}\hSlide{+2.0in}\ForceHeight{2.5in}\BoxedEPSF{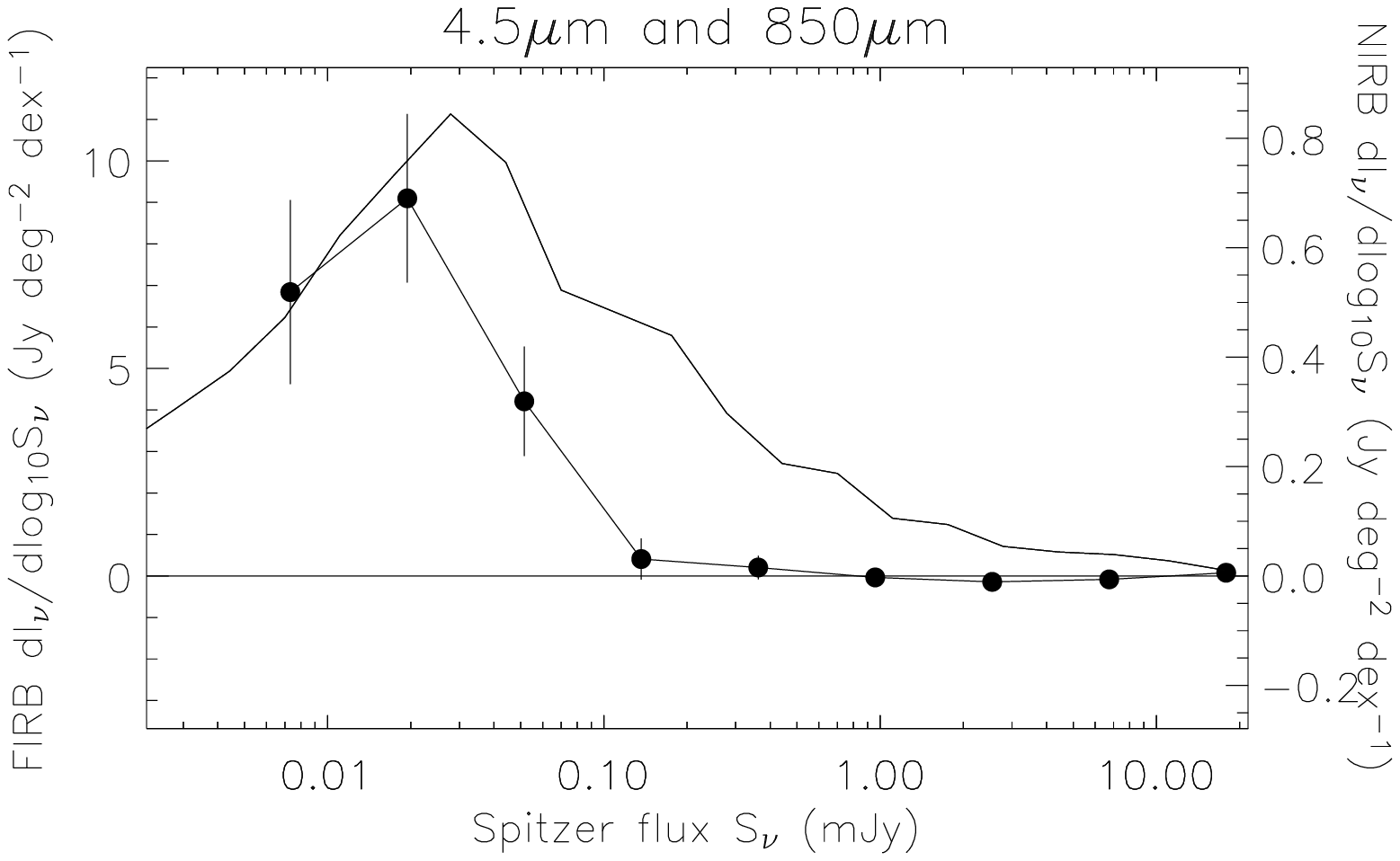}
                  \hSlide{-2.0in}\ForceHeight{2.5in}\BoxedEPSF{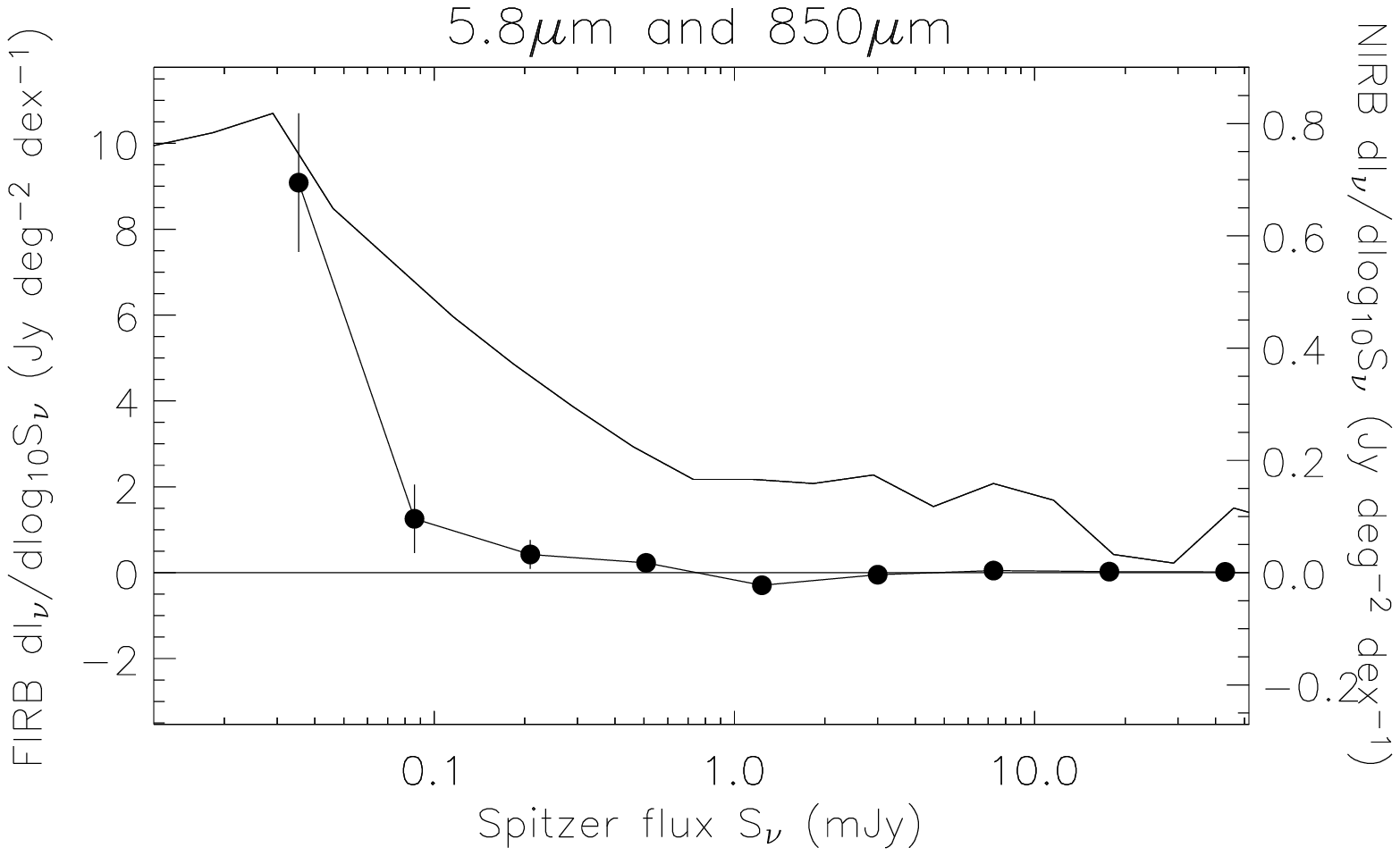}
  \vspace*{-2.5in}\hSlide{+2.0in}\ForceHeight{2.5in}\BoxedEPSF{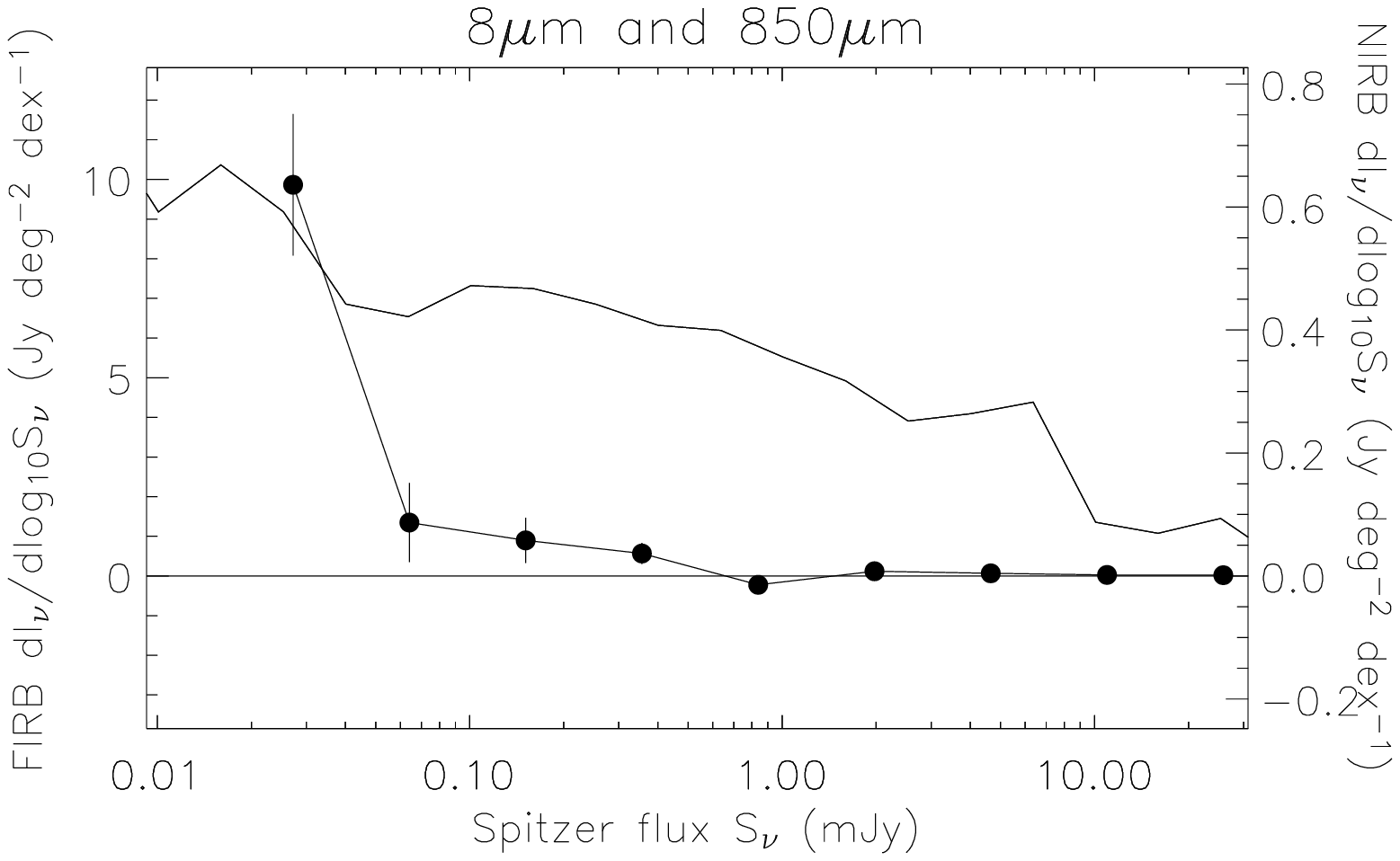}
                  \hSlide{-2.0in}\ForceHeight{2.5in}\BoxedEPSF{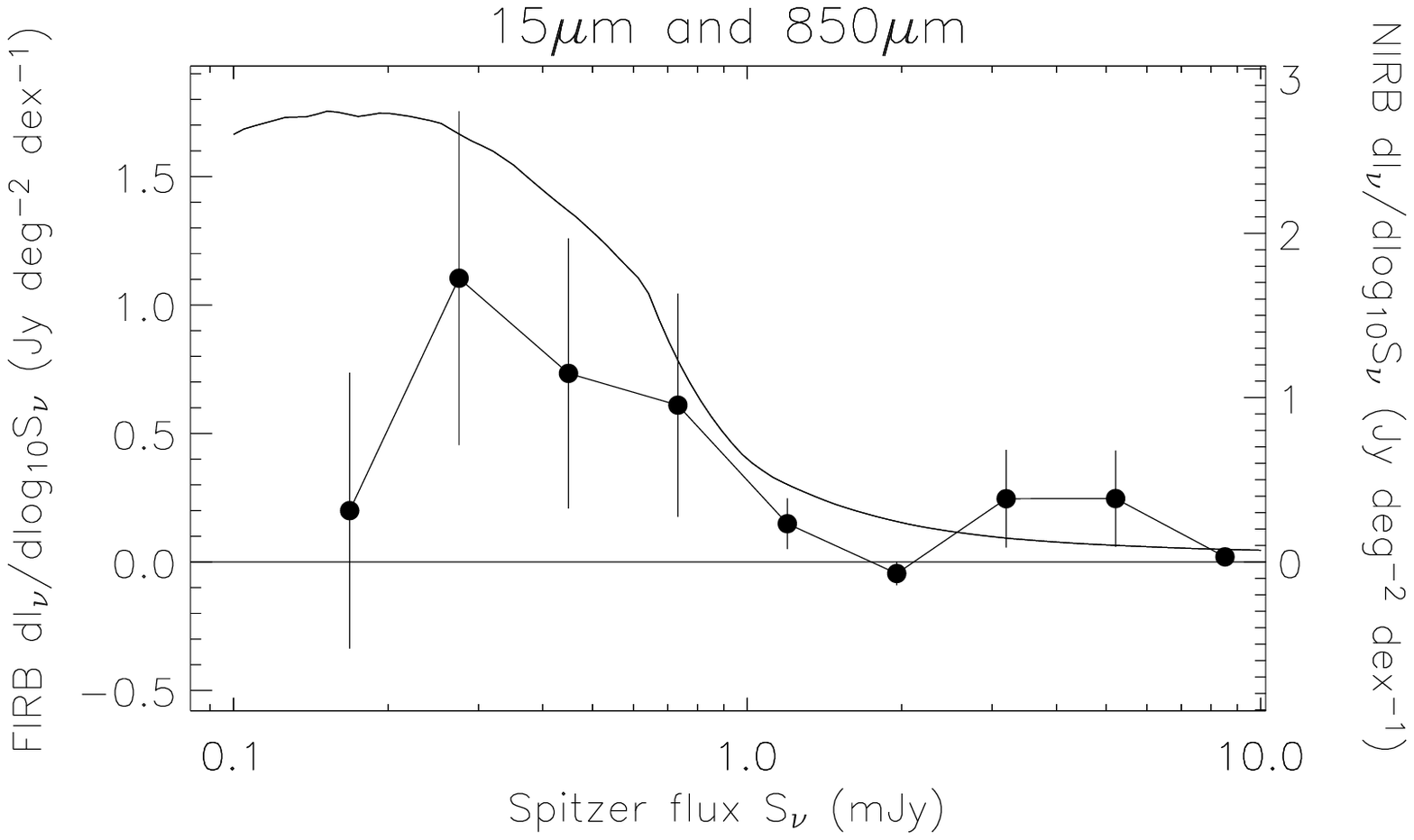}
  \vspace*{-2.5in}\hSlide{+2.0in}\ForceHeight{2.5in}\BoxedEPSF{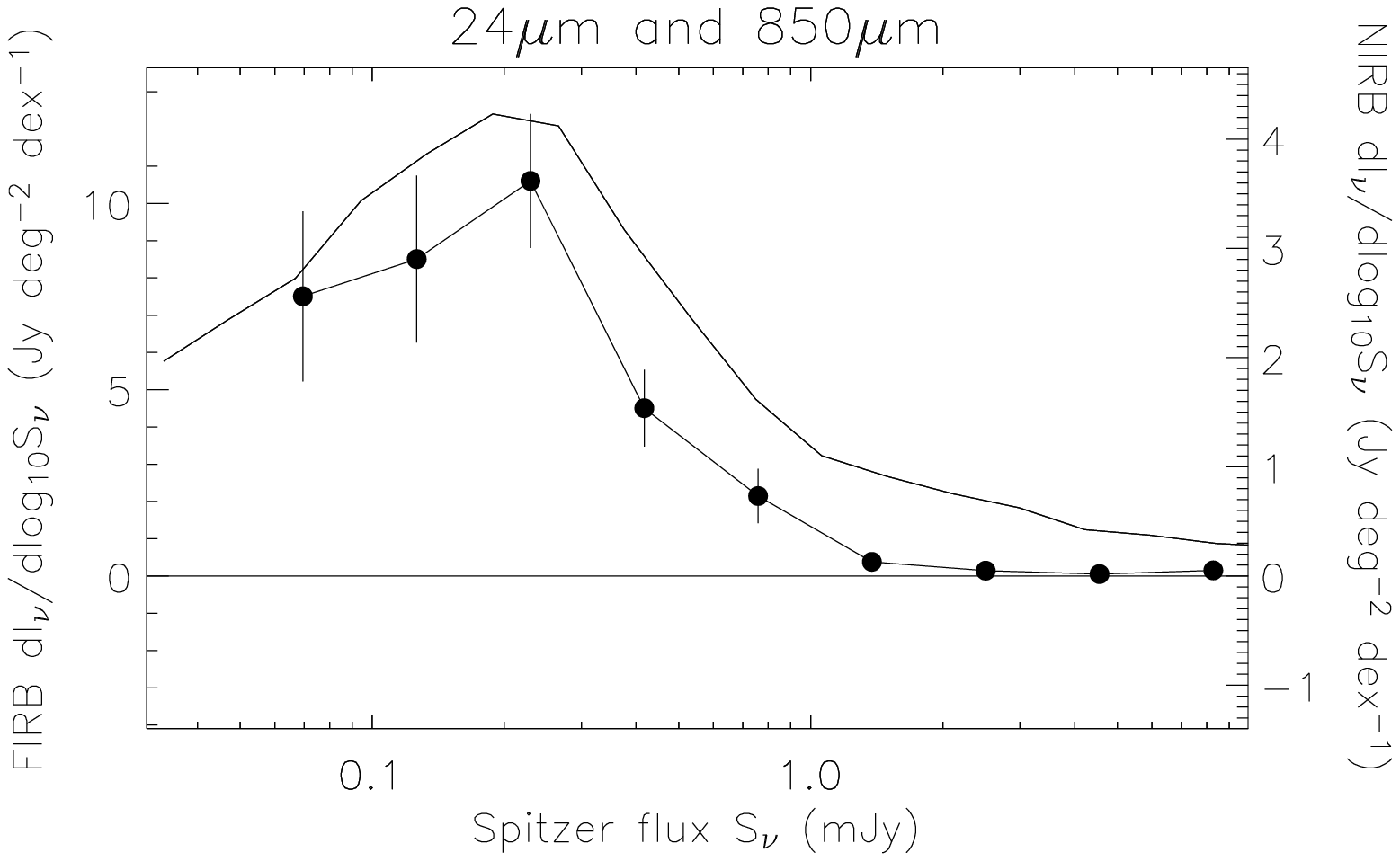}
\caption{\label{fig:850cfirb}Contributions of the Spitzer/ISO galaxies
  to the $850\,\mu$m extragalactic background light (data points,
  left-hand ordinates) as a function of Spitzer/ISO flux. Also plotted
  are the contributions of the same galaxies to the cosmic
  near-infrared and mid-infrared backgrounds (curves, right-hand
  ordinates), estimated from the source counts as discussed in the
  text. Note in particular the correspondence between the $24\,\mu$m and
  $850\,\mu$m extragalactic backgrounds.}
\end{figure*}

\begin{figure*}
                  \hSlide{-2.0in}\ForceHeight{2.5in}\BoxedEPSF{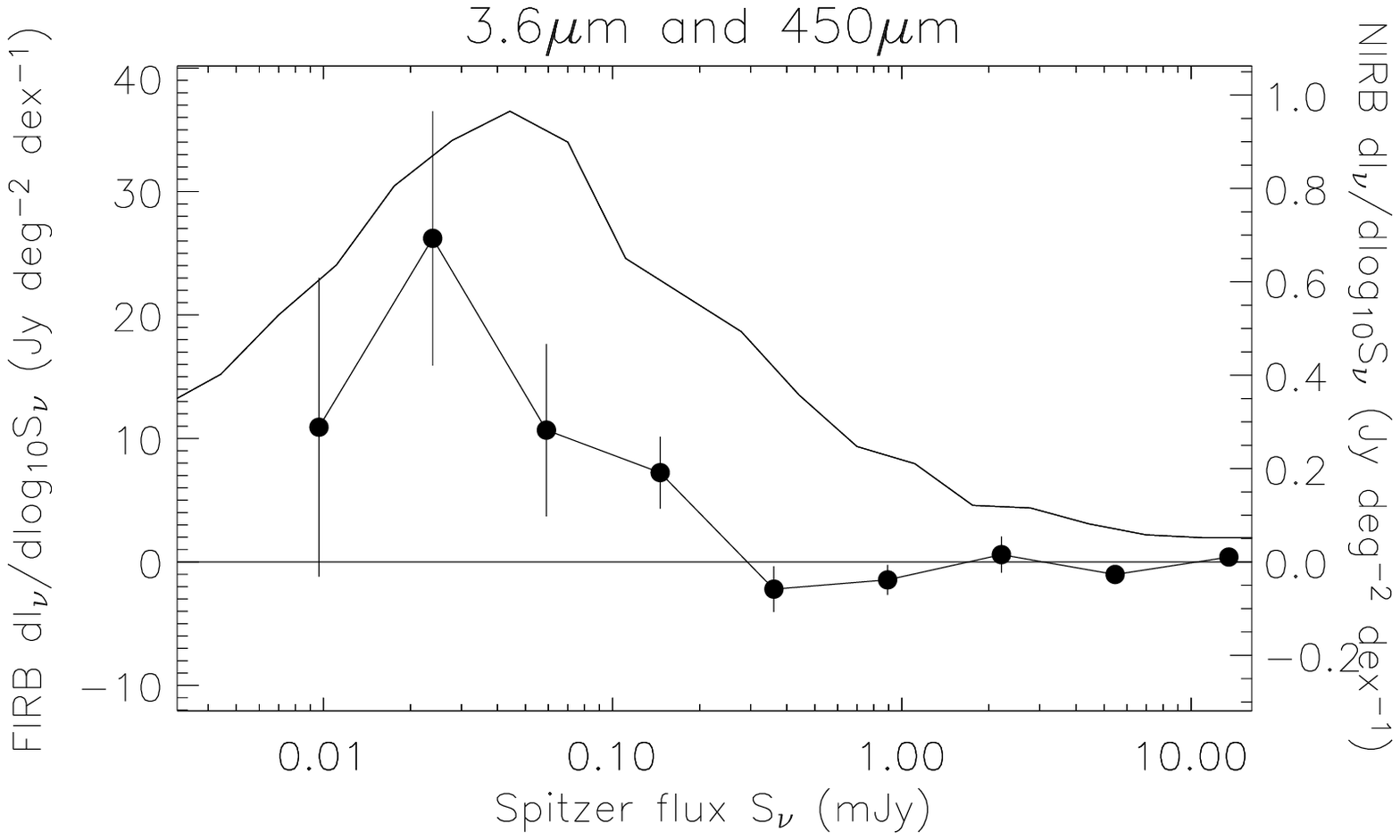}
  \vspace*{-2.5in}\hSlide{+2.0in}\ForceHeight{2.5in}\BoxedEPSF{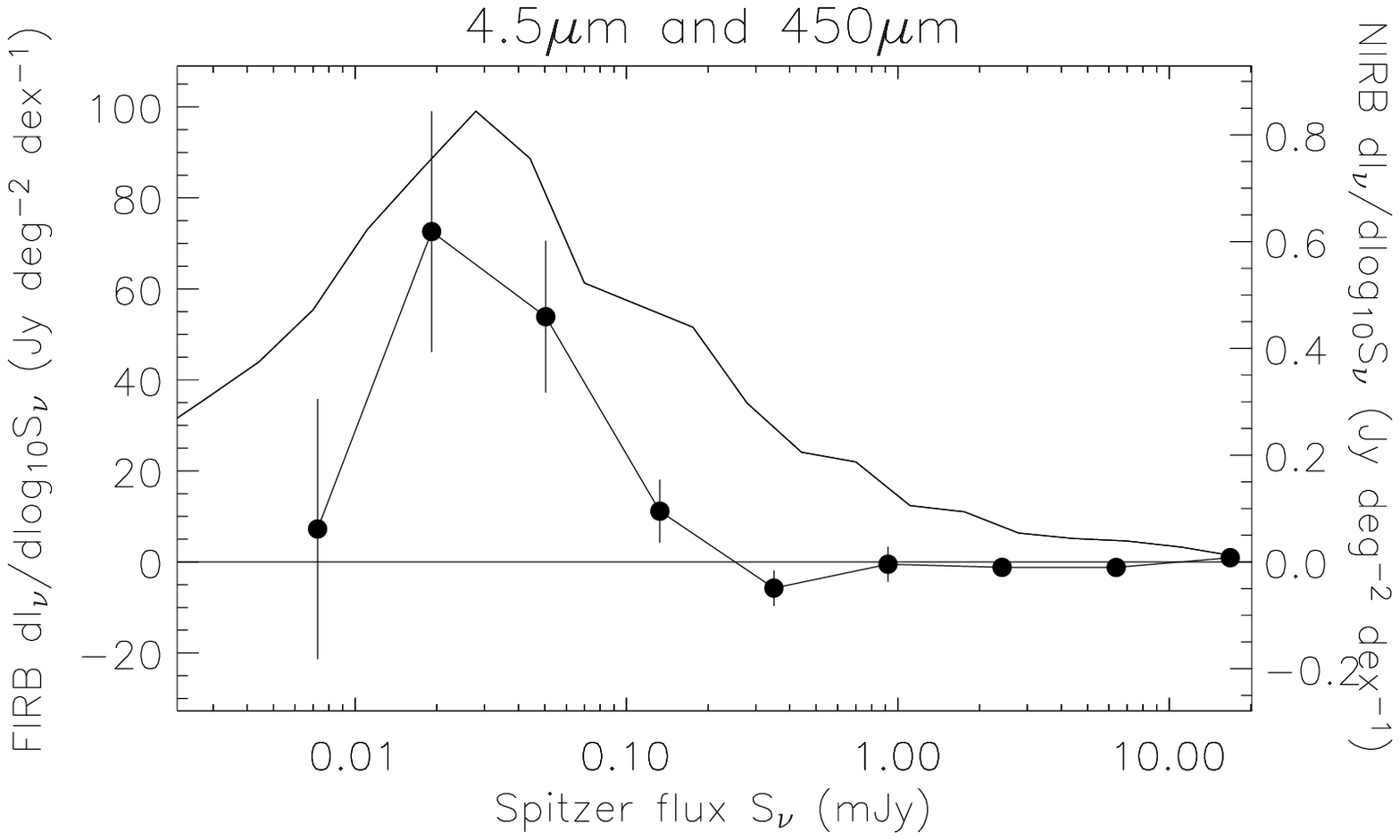}
                  \hSlide{-2.0in}\ForceHeight{2.5in}\BoxedEPSF{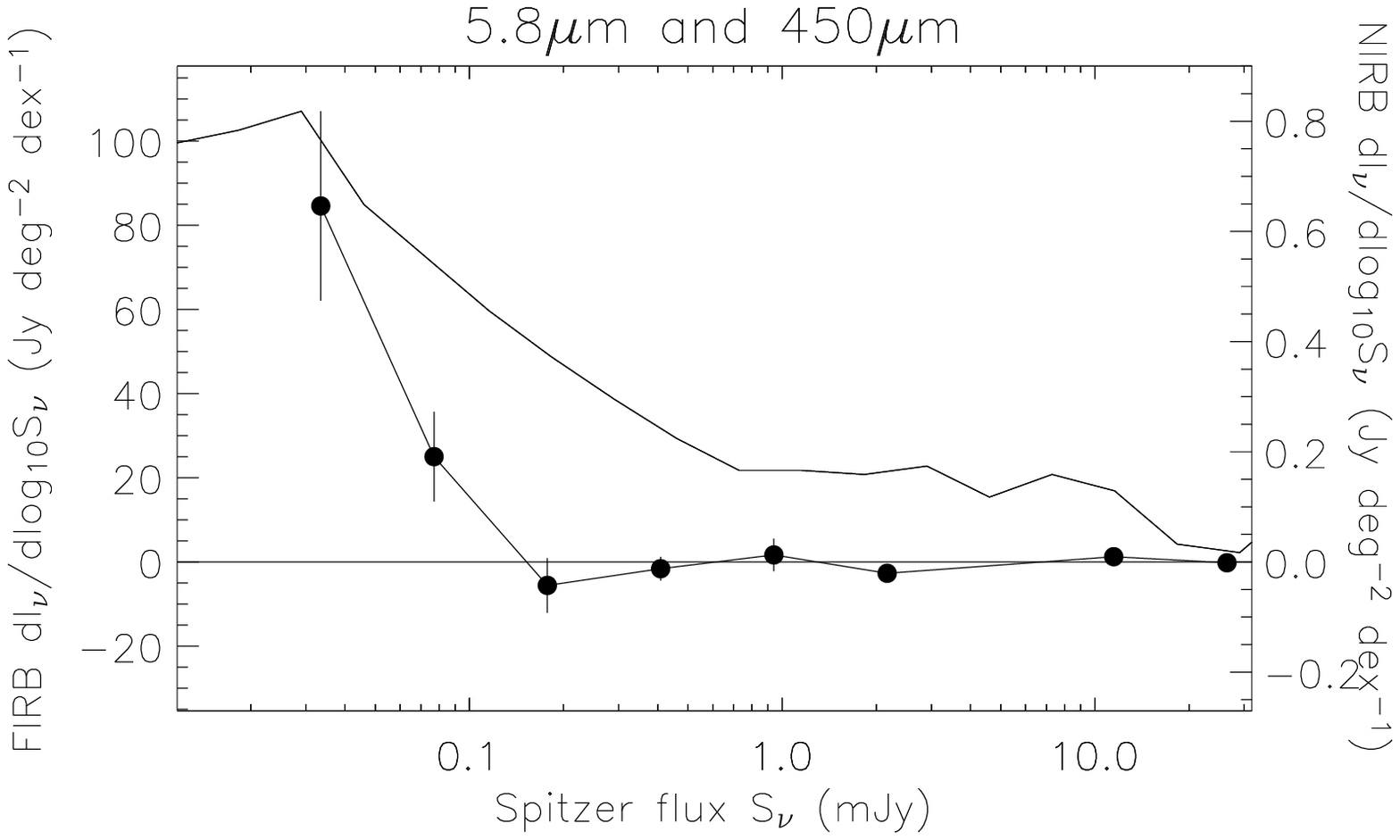}
  \vspace*{-2.5in}\hSlide{+2.0in}\ForceHeight{2.5in}\BoxedEPSF{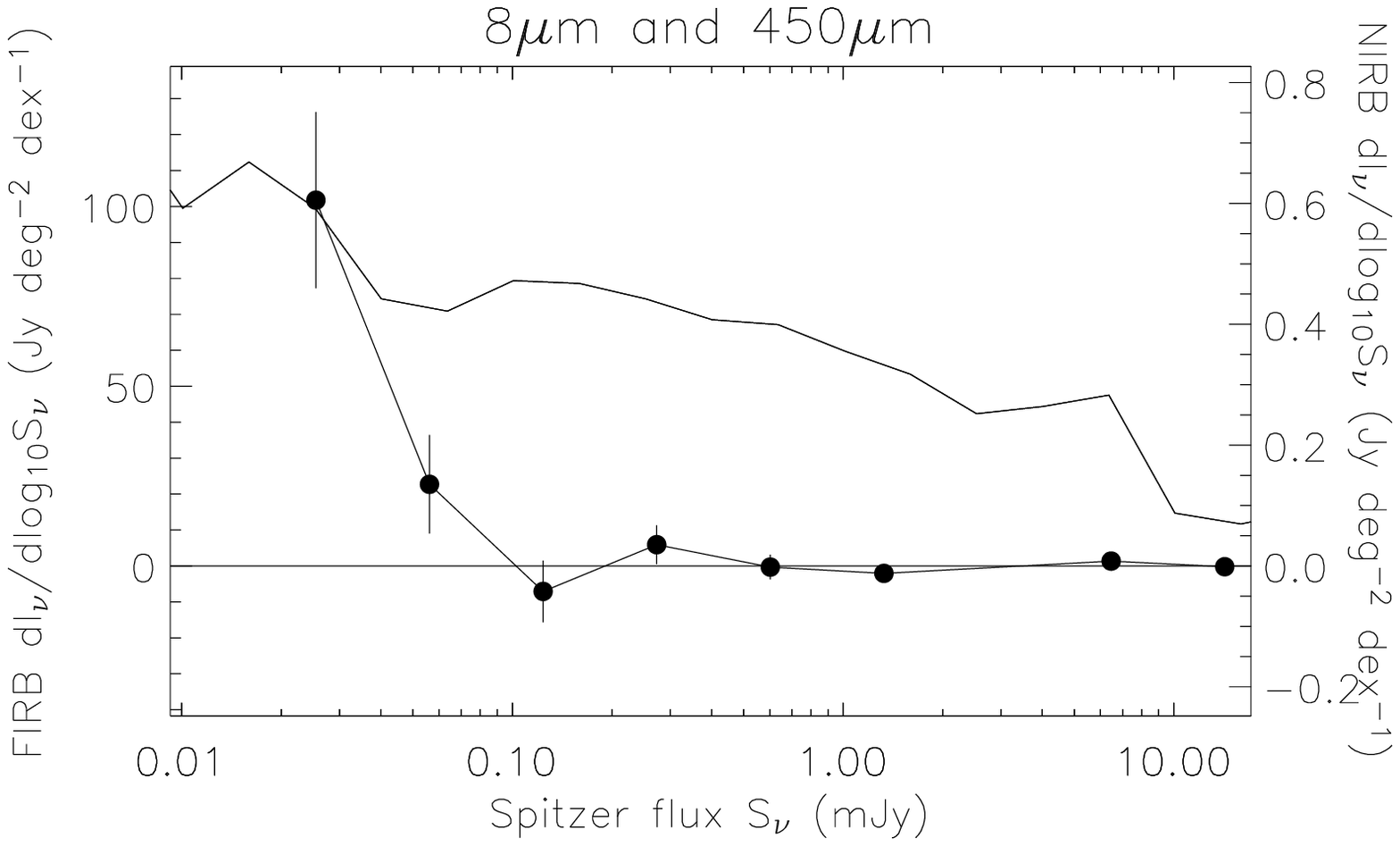}
                  \hSlide{-2.0in}\ForceHeight{2.5in}\BoxedEPSF{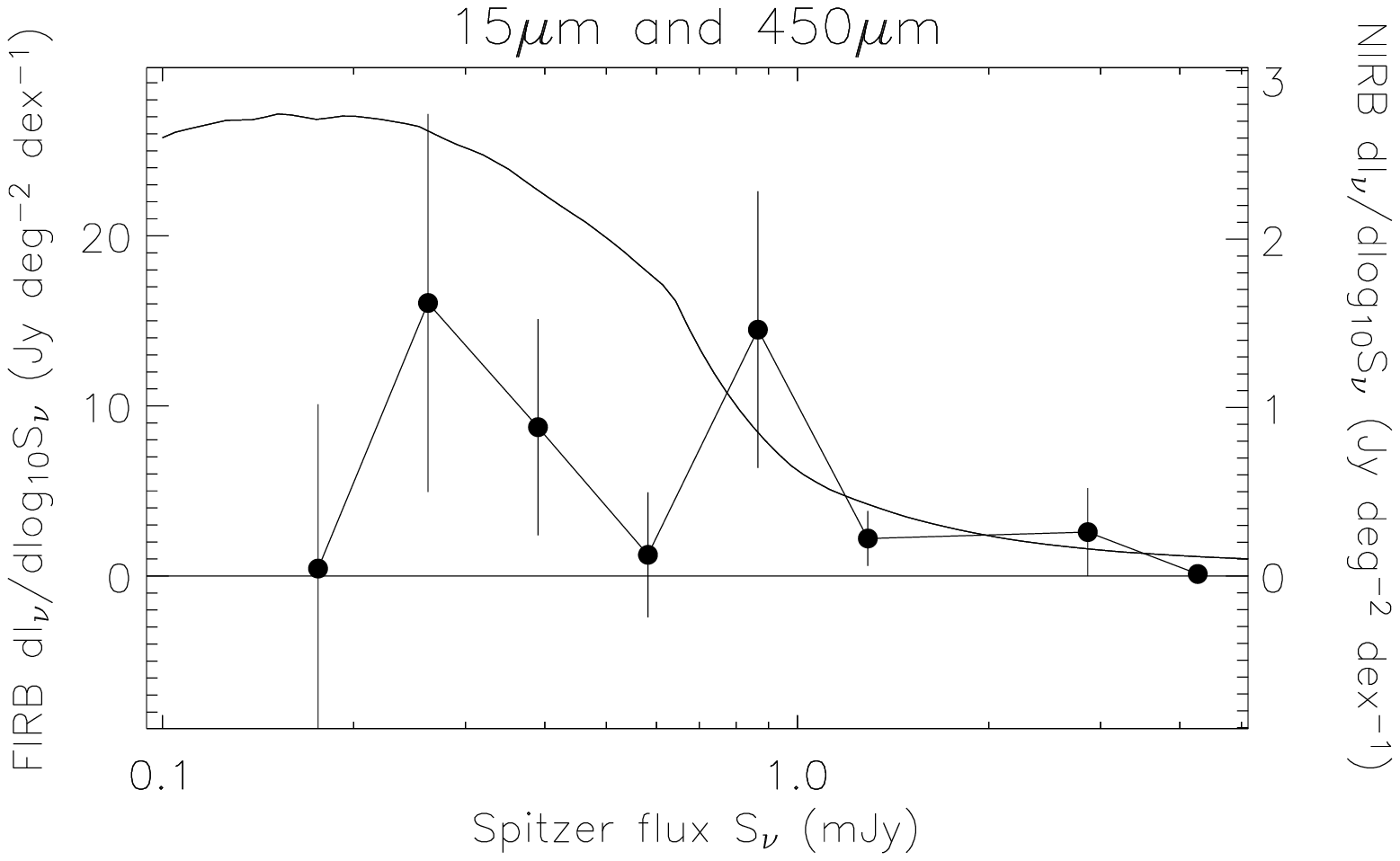}
  \vspace*{-2.5in}\hSlide{+2.0in}\ForceHeight{2.5in}\BoxedEPSF{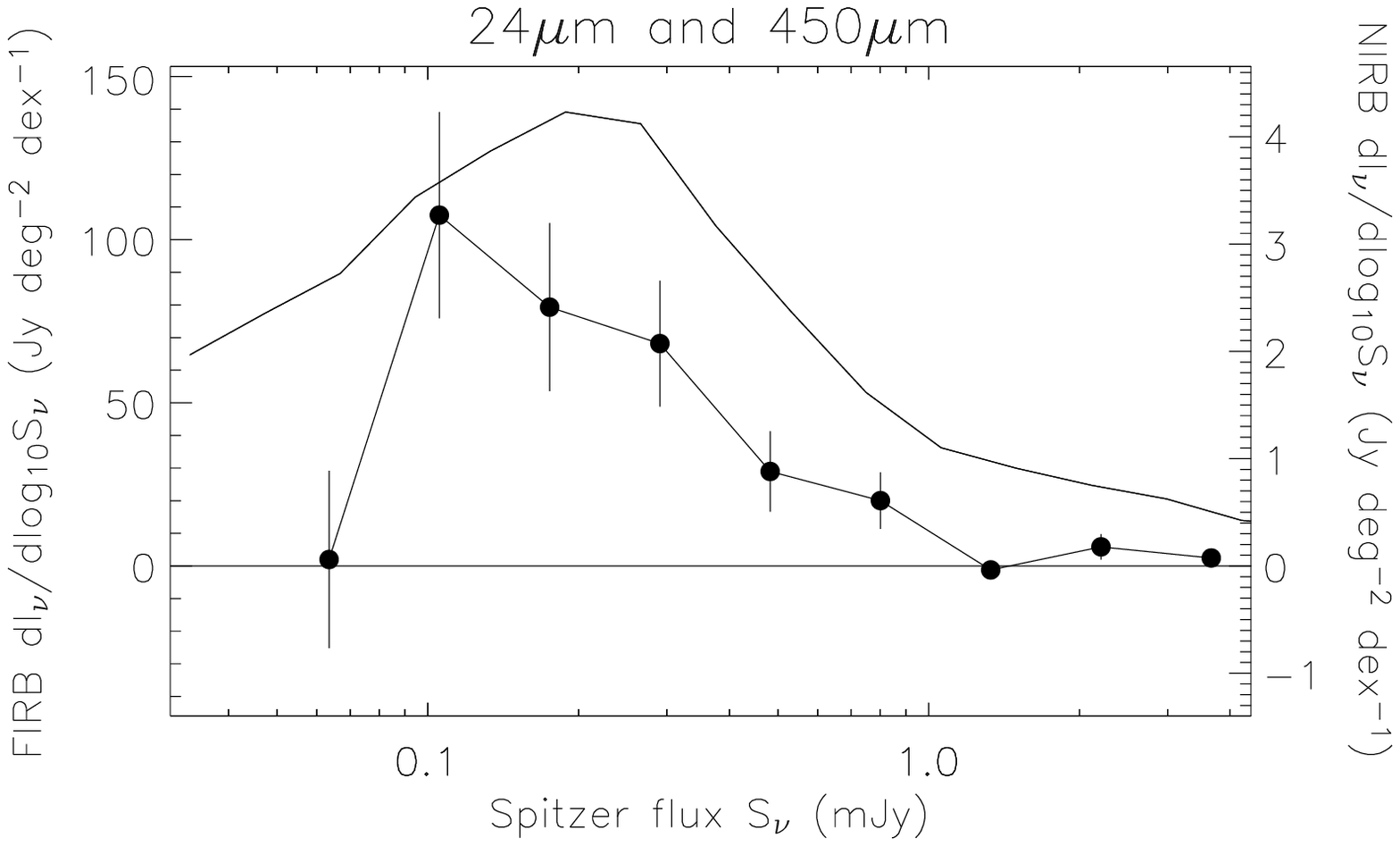}
\caption{\label{fig:450cfirb}Contributions of the Spitzer/ISO galaxies
  to the $450\,\mu$m extragalactic background light (data points,
  left-hand ordinates) as a function of Spitzer/ISO flux. Also plotted
  are the contributions of the same galaxies to the cosmic
  near-infrared and mid-infrared backgrounds (curves, right-hand
  ordinates), estimated from the source counts as discussed in the
  text. Note in particular the correspondence between the
  $3.6-4.5\,\mu$m and $450\,\mu$m extragalactic backgrounds.}
\end{figure*}

In figures \ref{fig:850cfirb} and \ref{fig:450cfirb} we plot the
contributions to the cosmic submm background light made by Spitzer/ISO
galaxies, as a function of their near/mid-infrared flux.
We correct
the Spitzer/ISO catalogues for incompleteness by comparison with
published source counts from Rodighiero et al. (2004), Papovich et
al. (2004), and Fazio et al. (2004). The figures show the submm
background contribution per decade of near/mid-infrared flux, and
compare these contributions to those made by the same galaxies to the
near/mid-infrared backgrounds calculated from the published source
counts. 

There is a remarkably strong correspondence between the
$24\,\mu$m-selected galaxy contribution to the $850\,\mu$m extragalactic
background, and to the $24\,\mu$m-selected background. There is also a
correspondence between the $3.6-4.5\,\mu$m-selected galaxy contributions
to the $450\,\mu$m extragalactic backgrounds, and to the $3.6-4.5\,\mu$m
backgrounds. The $24\,\mu$m background contributions also correlate
well with the $450\,\mu$m contributions. 
We will discuss the reasons for these correspondences in
section \ref{sec:discussion}. 

In figures 
\ref{fig:850cfirb_cumulative} and 
\ref{fig:450cfirb_cumulative} we integrate the data in figures
\ref{fig:850cfirb} and \ref{fig:450cfirb}, and plot the cumulative
contributions to the cosmic submm background light, as a function of
Spitzer/ISO flux. It is clear that about a quarter of the extragalactic
$850\,\mu$m background light is resolved by Spitzer, and the majority of
the $450\,\mu$m extragalactic background is resolved.

\subsection{The impact of clustering on stacking analyses}
\label{sec:method:spitzerclustering}
\begin{figure}
  \ForceWidth{3.5in}
  \BoxedEPSF{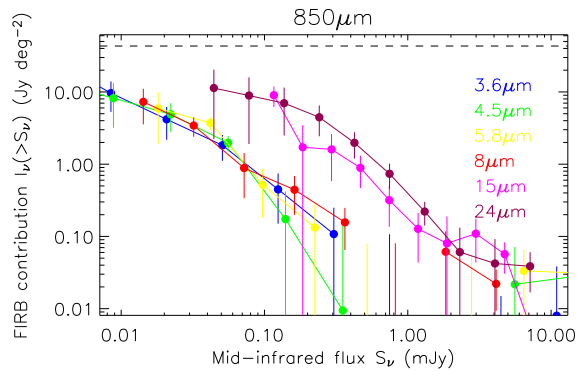}
\caption{\label{fig:850cfirb_cumulative} Cumumative contributions to
  the $850\,\mu$m extragalactic background light, as a function of
  near/mid-infrared flux, for various Spitzer/ISO surveys. The
  horizontal dashed line is the total background derived by Lagache et
  al. (1999), the uncertainties on which are less than a factor of
  $2$. Note that the Spitzer contributions converge to roughly a
  quarter of the total $850\,\mu$m background.}
\end{figure}

\begin{figure}
  \ForceWidth{3.5in}
  \BoxedEPSF{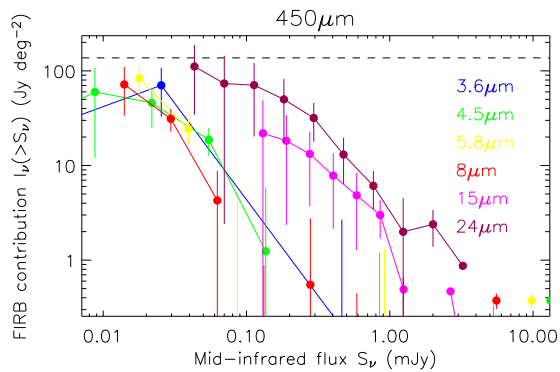}
\caption{\label{fig:450cfirb_cumulative} Cumumative contributions to the
  $450\,\mu$m extragalactic background light, as a function of
  near/mid-infrared flux, for various Spitzer/ISO surveys. Symbols as
  figure \ref{fig:850cfirb_cumulative}. Note that the Spitzer
  contributions are capable of accounting for the entire $450\,\mu$m
  background, within the errors.}
\end{figure}

Because of the limited resolution provided by SCUBA, we should be
clear that the signal detected in the stacking analysis represents a
contribution from any object within around $10''$ of the target
Spitzer/ISO galaxies. Because these galaxies will be surrounded by a
population of correlated neighbours, it is therefore possible that the
stacked flux gives an overestimate of the emission from the target
galaxies. This is simple enough to estimate: the additional flux is
just the integral of the background intensity, $I$, times the angular
cross-correlation between the target galaxies and the background,
$w(\theta)$, times the beam $B$:
\begin{equation}\label{eqn:wtheta}
S = I \int w(\theta)\,B(\theta)\,2\pi\theta\,d\theta
\end{equation}
For a Gaussian beam, and assuming $w=(\theta/\theta_0)^{-0.8}$, this
gives $S=0.20\,(\theta_0/{\rm arcsec})^{0.8}$\,mJy at $850\,\mu$m and
$S=0.28\,(\theta_0/{\rm arcsec})^{0.8}$\,mJy at $450\,\mu$m. The
appropriate value of $\theta_0$ is of course open to debate, but
Oliver et al. (2004) measure $\theta_0=1.24''$ for the
$3.6\,\mu$m-selected population. The stacked fluxes at $450\,\mu$m are
at least five times the maximum value that could arise from 
neighbours (taking the extreme case in which all the $850\,\mu$m
signal arises in this way). At $850\,\mu$m the contribution may be
more significant, but the contribution estimated from equation
\ref{eqn:wtheta} is necessarily an over-estimate because not all the
$850\,\mu$m background is attributable to these Spitzer galaxies.  We
can use figure \ref{fig:850cfirb_cumulative}, in which only
$\simeq25\%$ of the $850\,\mu$m background is attributable to the
Spitzer galaxies, to estimate iteratively the correlated $850\,\mu$m
flux from the other Spitzer galaxies. This reduces the
clustered contribution by a factor of four, yielding $S=0.06\,$mJy
with $\theta_0=1.24''$, which is at most a $20-30\%$ correction to the
$850\,\mu$m fluxes quoted in table \ref{tab:meanfluxes}. While
non-zero, this is not sufficient to affect our conclusions.

\section{Discussion}\label{sec:discussion}

\subsection{The link between the near-IR and submm backgrounds}
Any stacking analysis is only capable of determining the first moment
of the distribution; the mean values in figures \ref{fig:madau} and
\ref{fig:specific_sfr} may belie a large variation in the population.
We have also subtracted point sources, so strongly starbursting galaxies are
omitted from these figures; we are therefore probing only the {\it
  mean quiescent} levels of star formation in these galaxies.

Our stacking detections have much higher signal-to-noise than any
previously obtained, partly because of the depths of the Spitzer and
SCUBA surveys, and partly also because the stacking signal-to-noise
scales with the square root of the number of submm beams and SHADES
has the largest contiguous submm survey fields to date. Our
$450\,\mu$m stacking results are the best indicators of the
populations that will be found to dominate the $450\,\mu$m background
by SCUBA-2. The prospects are good for follow-ups of the ultradeep
SCUBA-2 Cosmology Survey, because the Spitzer galaxies that appear to
dominate the $450\,\mu$m background are less challenging targets for
$8-10$m-class spectroscopy than SCUBA point sources (e.g. Serjeant et
al. in preparation). Similarly, the prospects appear good for
spectroscopic follow-ups of ALMA point sources below the SCUBA-2
confusion limit. The clustering of bright submm point sources as a
function of redshift is a key goal of the SHADES survey (e.g. van
Kampen et al. 2005, Mortier et al. 2006), providing strong constraints
on semi-analytic models of galaxy evolution; similarly, the
redshift-dependent clustering of galaxies a factor of $\sim10$ fainter
in submm flux, for which optical follow-up is easier, is likely also
to provide a strong constraint on such models.

Paradoxically, it is the brighter submm point sources that are the
most challenging to follow up.  The populations sampled by the SPIRE
instrument on Herschel will be challenging to follow up in the optical
(see also e.g. Khan et al. 2005, 2007). Spectroscopic redshifts for
such populations may be better determined in the medium term by
molecular line spectroscopy (e.g. Wagg et al. 2007), and in the longer
term by SPICA (e.g. Nakagawa 2004)

We find that the submm:$24\,\mu$m ratios for most $24\,\mu$m-selected
galaxies are very different to those of most submm-selected galaxies,
in agreement with Serjeant et al. (2004). If the $24\,\mu$m population
had submm:$24\,\mu$m flux ratios consistent with those of the submm
point source population, the $24\,\mu$m sources considered in this
paper would overproduce the $850\,\mu$m background by a factor of
$\sim\times3$.  We argue that the bulk of the Spitzer population has a
quiescent star formation level much lower than that of submm point
sources, while the latter are heavily obscured objects (e.g. Serjeant
et al. 2003b, Clements et al. 2004, Smail et al. 2004) that are
challenging for optical follow-ups, and far-infrared ``loud'' with
high specific star formation rates (figure
\ref{fig:shades_specific_sfr}) and short star formation
timescales. Such episodic star formation is supported by models of AGN
feedback in massive galaxies in the early Universe, and by the small
inferred mass accretion rates onto central supermassive black holes in
submm point sources (e.g. Alexander et al. 2005).

The physical sizes of populations dominating the far-IR background are
rather smaller than those inferred for submm point sources. At a flux
density of $\sim15\mu$Jy, the $8\,\mu$m population which we have found
to contribute significantly to the $450\,\mu$m extragalactic
background has optical identifications in our imaging with typical
diameters $\stackrel{<}{_\sim}1''$ (the optical identification 
diameters of submm point sources can be up to $\sim2-4''$, e.g. Smail et
al. 2004, Pope et al. 2005, though other authors claim sub-arcsecond
sizes, e.g. Chapman et al. 2004, 
Biggs \& Ivison 2008).  The regions of star formation in these
galaxies will be resolvable with ALMA, which will probe the cool
large-grained dust phase, and Darwin direct imaging which will probe
the transiently-heated small grains and PAH phases. This also suggests
that $0.1''$ is the coarsest resolution that would be useful for a
future $\sim50-200\,\mu$m far-infrared interferometer (FIRI) to
resolve the internal structure of individual galaxies that comprise
the cosmic far-infrared background.

\subsection{The mass dependence of star formation}\label{sec:discussion:mass}
When plotting the total contribution to the submm background from
redshift shells in our Spitzer samples (not shown), we find that the
$z\stackrel{<}{_\sim}1.5$ population is dominant, similar to the
results of Wang et al. (2006). However, this
neglects the fact that different luminosity and mass ranges are
sampled at different redshifts. This may be one underlying cause of
the difference between the Wang et al. (2006) and Dye et al. (2006)
stacking results, since the K-correction effects are different in
their respective samples; cosmic variance is another possibility. Ours
is the first direct attempt to segregate the mass contributions to the
submm-derived cosmic star formation history.

The mass segregation in figure \ref{fig:madau} shows evidence for star
formation in galaxies with model stellar masses $\simeq10^{10}M_\odot$
assembling the bulk of their stellar masses at much lower redshifts
than the estimated $z\sim2.2$ peak in the submm point source
population.
This implies an increasing dominance at higher redshift of higher mass
systems in the volume-averaged star-formation rate. These observations
are in accordance with qualitative expectations from ``downsizing''
in star formation (Matteucci 1994, Bressan et al. 1996, Cowie et
al. 1996; see also papers VII and VIII, Dye et al. 2008 and Clements
et al. 2007 respectively). 

Figure \ref{fig:mean_sfr} shows the mass- and redshift-dependence of
the mean star formation rates per {\it galaxy}, rather than per unit
volume. The decrease in the mean star formation rates per galaxy at
$\sim10^{10}\,M_\odot$ and $z<1$ can only be reconciled with their
increasing volume-averaged contribution if their number densities are
increasing, in agreement with expectations from mass downsizing
(e.g. Pozzetti et al. 2007). Figure \ref{fig:mean_sfr}
also shows the corresponding
predictions from the de Lucia (2005) Millenium simulation. Our star
formation rates are a factor of a few higher than those predicted in
this simulation.

The star formation timescales in figure \ref{fig:specific_sfr} at
redshifts \mbox{$z>1$} scale approximately inversely with the mass of
the system, quite unlike e.g. a Schmidt law for local late-type
galaxies.  In figure \ref{fig:specific_sfr} we again compare this to
the predictions from the de Lucia (2005) Millenium simulations. Again,
there is a striking discrepancy with the simulation predictions.

How bad would the photometric redshifts have to be in order to explain
the disagreement between model and data? The results in figure
\ref{fig:specific_sfr} are surprisingly insensitive to photometric
redshift, and it is therefore not likely that errors in the
photometric redshifts are the cause of the disagreement. This
insensitivity to redshift errors is due to the model stellar mass and
star formation rates both being relatively insensitive to redshift
(e.g. figure \ref{fig:mstar_vs_z}) because both are to some degree
subject to negative K-corrections. The specific star formation rate
estimates therefore depend mainly on the submm:near-IR flux ratio, and
not on redshift. It is therefore hard to see how a redistribution of
galaxies among the redshift bins could bring the data into agreement
with the models; removing galaxies from one redshift bin to make it
agree better with the models would make the disagreement worse in the
other bins into which the galaxies are moved.

One hint that the star formation rates may be overestimated comes from
asking what will happen to the $10^{10}M_\odot$ galaxies if they
continue forming stars at these high continuous rates. The galaxies
appear to be forming stars at over ten times the rate required to
double their mass by the present day, almost regardless of redshift at
$z<2$. This is clearly not sustainable; 
it would require a rapid truncation of the star formation at
slightly higher masses, and a continual feeding of lower mass galaxies
into the $\sim 10^{10}M_\odot$ bin, for which there is no evidence in
the stellar mass functions of galaxies (e.g. Bell et al. 2004, 2007,
Caputi et al. 2006).

\begin{figure}
  \ForceWidth{3.5in}
  \BoxedEPSF{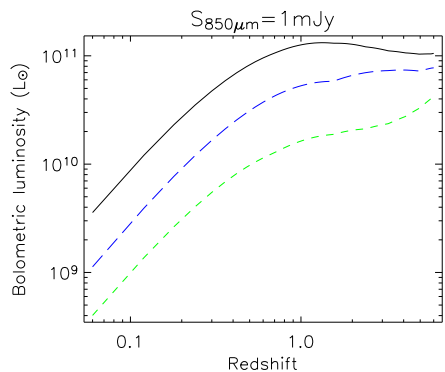}
\caption{\label{fig:lbol} 
  Bolometric luminosity as a function of redshift for an observed 
  $850\,\mu$m flux of $1$\,mJy, for three spectral energy distributions: 
  M82 (full line), a cirrus-dominated spectrum from Efstathiou et al. 2000 
  (short dashed, green), 
  and an Arp 220 model also from Efstathiou et al. 2000 (long dashed, green).} 
\end{figure}

There are at least two possibilities that might reduce our estimates
of the number of stars forming in these galaxies. 
One approach
explored successfully by the Baugh et al. (2005) application of the
Millenium simulation is to assume a top-heavy initial mass function in
star-forming galaxies. The reduction in the number of stars formed in
this model is similar to the discrepancy between the de Lucia (2005)
predictions and our measurements. However, the top-heavy initial mass
function is normally only applied to extreme starbursting systems, and
not to the quiescent star formation level in the galaxy population as
a whole. A second possibility is that the observed-frame submm fluxes
are dominated not by star formation, but by cool cirrus heated by the
galaxies' interstellar radiation fields. The galaxies would then have
a cooler SED than the M82 template assumed above. We demonstrate the
strength of this effect in figure \ref{fig:lbol}. Such a model was
proposed for bright submm point sources by Efstathiou \&
Rowan-Robinson 2003 (see also Clements et al. 2007, paper VIII), and
although mm-wave and radio interferometry 
have not on the whole yielded the large angular sizes predicted by
these models (e.g. Tacconi et al. 2006, Ivison et al. 2007), it
remains possible that these models are broadly correct descriptions of
the fainter submm population.

\subsection{The environment dependence of star formation}
We have attempted to measure the matter environments of submm
galaxies. Semi-analytic models predict that submm galaxies should be
strongly clustered and lie in some of the largest overdensities at
their redshifts (e.g. van Kampen et al. 2005), for which we have found
tentative evidence (e.g. figures \ref{fig:percentiles} and
\ref{fig:percentiles_2Mpc}) and which agrees at least qualitatively
with previous measurements (e.g. Blain et al. 2004, 
Blake et al. 2006). Recently, Elbaz
et al. (2007) and Cooper et al. (2007) have shown evidence in the
GOODS and DEEP2 surveys that $z\sim1$ star-forming galaxies (as
evidenced by $24\mu$m emission) are preferentially found in richer
galaxy environments. This trend is in the opposite sense to that seen
in the local Universe. Furthermore, the observed environment
dependence of star formation is stronger than that predicted by
semi-analytic models. Our observations extend this trend to higher
redshifts and star formation rates. The properties of the {\it star
  formation} density field, as opposed to the galaxy density field,
may be a key arena for the future confrontation of data and
semi-analytic predictions. Future Herschel and SCUBA-2 surveys will
yield large catalogues of bright submm-selected galaxies, and their
near-infrared and submm environments will be easily measurable with
the warm AKARI/Spitzer missions and SCUBA-2 follow-ups respectively.

\section{Conclusions}\label{sec:conclusions}
There is a strong correspondence between the galaxies that dominate
the submm extragalactic background light and those that are detected
in deep Spitzer surveys. The submm-derived specific star formation
rates in the Spitzer populations are much higher than those predicted
by some semi-analytic simulations; this may be due to a component of
submm emission heated by the interstellar radiation fields leading to
overestimates of the star formation rates, or to a top-heavy initial
mass function in the Spitzer galaxies, or to some unknown deficiency
in the models. We find evidence for downsizing in both star formation
and mass assembly.  We also find evidence that around a third of
submm-selected galaxies at redshifts $1<z<1.5$ lie in the upper
$\sim20\%$-ile of the galaxy density distribution, in contrast to the
redshift zero tendency of star-forming galaxies to avoid the richest
environments.

\section*{Acknowledgements}
We would like to thank the referee, Wei-Hao Wang, for a careful
reading of the manuscript and for helpful comments. 
SS would like to thank the Science and Technology Facilities Council
for support under grants PP/D002400/1 and PP/D003083/1. KC and SD
would like to thank the Science and Technology Facilities Council for
support.  
IRS acknowledges support from the Royal Society. 
AP acknowledges support provided by NASA through the Spitzer Space
Telescope Fellowship Program, through a contract issued by the Jet
Propulsion Laboratory, California Institute of Technology under a
contract with NASA.
Data was obtained at the James Clerk Maxwell Telescope under
program M/02B/U52. The James Clerk Maxwell Telescope is operated by
The Joint Astronomy Centre on behalf of the Science and Technology
Facilities Council of the United Kingdom, the Netherlands Organisation
for Scientific Research, and the National Research Council of Canada.
This work is based in part on observations made with the Spitzer Space
Telescope, which is operated by the Jet Propulsion Laboratory,
California Institute of Technology under a contract with NASA.

\end{document}